\documentclass[11pt]{article}

\parskip \medskipamount
\usepackage{amsfonts}
\usepackage[round]{natbib}	
\newcommand*{\doi}[1]{\href{http://dx.doi.org/#1}{doi: #1}}
\usepackage{graphicx} 									
\usepackage[small,bf,skip = 3pt]{caption}	
\usepackage{subcaption}		
\usepackage{wrapfig}										
\usepackage{multirow}
\usepackage[table]{xcolor}										
\usepackage[latin1]{inputenc}						
\usepackage[a4paper, left=2.5cm, right=2.5cm, top=1.8cm, bottom=1.8cm, includehead, includefoot, head=30pt]{geometry} 	
\usepackage{amsmath}
\allowdisplaybreaks[1]									
\usepackage{amssymb}										
\usepackage{amsthm}											
\usepackage{booktabs}										
\usepackage{varwidth}
\usepackage{url}                        
\usepackage{eurosym}                    
\usepackage{textcomp}                   
\usepackage{fancyhdr}                   
\usepackage{listings} 									
\usepackage{pdfpages}										
\usepackage{float}                      
\usepackage[title,titletoc,toc]{appendix}				
\usepackage{enumerate}									
\usepackage[multiple]{footmisc}																	
\usepackage{enumitem}										
\usepackage{mathtools}										
\usepackage{IEEEtrantools}							
\usepackage[ruled]{algorithm2e} 
\SetAlCapFnt{\small} 
\SetKw{KwTo}{in}
\usepackage{authblk}											
\usepackage{tikz}												
\usetikzlibrary{arrows,decorations.pathmorphing,backgrounds,positioning,fit,petri}
\usepackage{rotating}
\usepackage{multirow}
\usepackage{grffile}
\usepackage{amsfonts}
\usepackage{amssymb}
\usepackage{rotating}
\usepackage{siunitx}
\usepackage{color}	
\usepackage{etoolbox}
\usepackage[plainpages=false,hyperfootnotes=false]{hyperref}
\hypersetup{
  colorlinks   = true, 
  urlcolor     = blue, 
  linkcolor    = red, 
  citecolor   = blue 
}
\usepackage{xurl}
\PassOptionsToPackage{hyphens}{url}\usepackage{hyperref}

\makeatletter

\pretocmd{\NAT@citex}{%
  \let\NAT@hyper@\NAT@hyper@citex
  \def\NAT@postnote{#2}%
  \setcounter{NAT@total@cites}{0}%
  \setcounter{NAT@count@cites}{0}%
  \forcsvlist{\stepcounter{NAT@total@cites}\@gobble}{#3}}{}{}
\newcounter{NAT@total@cites}
\newcounter{NAT@count@cites}
\def\NAT@postnote{}

\def\NAT@hyper@citex#1{%
  \stepcounter{NAT@count@cites}%
  \hyper@natlinkstart{\@citeb\@extra@b@citeb}#1%
  \ifnumequal{\value{NAT@count@cites}}{\value{NAT@total@cites}}
    {\ifNAT@swa\else\if*\NAT@postnote*\else%
     \NAT@cmt\NAT@postnote\global\def\NAT@postnote{}\fi\fi}{}%
  \ifNAT@swa\else\if\relax\NAT@date\relax
  \else\NAT@@close\global\let\NAT@nm\@empty\fi\fi
  \hyper@natlinkend}
\renewcommand\hyper@natlinkbreak[2]{#1}

\patchcmd{\NAT@citex}
  {\ifNAT@swa\else\if*#2*\else\NAT@cmt#2\fi
   \if\relax\NAT@date\relax\else\NAT@@close\fi\fi}{}{}{}
\patchcmd{\NAT@citex}
  {\if\relax\NAT@date\relax\NAT@def@citea\else\NAT@def@citea@close\fi}
  {\if\relax\NAT@date\relax\NAT@def@citea\else\NAT@def@citea@space\fi}{}{}

\makeatother

\definecolor{darkgreen}{RGB}{40,150,40}

\newcommand{\bucket}[1]{#1_{[x_i,x_{j}],t,w}}

\renewcommand{\geq}{\geqslant}					
\renewcommand{\epsilon}{\varepsilon}

\theoremstyle{definition}

\usepackage{adjustbox}

\usepackage{tablefootnote}

\definecolor{light-gray}{gray}{0.90}

\author[]{Jens Robben\footnote{Corresponding author. E-mail address: \href{mailto:jens.robben@kuleuven.be}{jens.robben@kuleuven.be}.}}
\author[1,2]{Katrien Antonio\footnote{Corresponding author. E-mail address: \href{mailto:katrien.antonio@kuleuven.be}{katrien.antonio@kuleuven.be}.}}
\author[]{Sander Devriendt\footnote{Voluntary research assistant on the mortality project at KU Leuven, AFI Department, Insurance Research Group. This paper reflects the personal views of the author and not the views of his employer.}}
\affil[1]{Faculty of Economics and Business, KU Leuven, Belgium.}
\affil[2]{Faculty of Economics and Business, University of Amsterdam, The Netherlands.}
\title{\textbf{Assessing the impact of the COVID-19 shock on a stochastic multi-population mortality model}}

\begin{document}
\sloppy
\maketitle
\interfootnotelinepenalty=10000
\begin{abstract}

We aim to assess the impact of a pandemic data point on the calibration of a stochastic multi-population mortality projection model and its resulting projections for future mortality rates. Throughout the paper we put focus on the Li \& Lee mortality model, which has become a standard for projecting mortality in Belgium and the Netherlands. We calibrate this mortality model on annual deaths and exposures at the level of individual ages. This type of mortality data is typically collected, produced and reported with a significant delay of - for some countries - several years on a platform such as the Human Mortality Database. To enable a timely evaluation of the impact of a pandemic data point we have to rely on other data sources (e.g.~the Short-Term Mortality Fluctuations Data series) that swiftly publish weekly mortality data collected in age buckets. To be compliant with the design and calibration strategy of the Li \& Lee model, we have to transform the weekly mortality data collected in age buckets to yearly, age-specific observations. Therefore, our paper constructs a protocol to ungroup the deaths and exposures registered in age buckets to individual ages. To evaluate the impact of a pandemic shock, like COVID-19 in the year 2020, we weigh this data point in either the calibration or projection step. Obviously, the more weight we place on this data point, the more impact we observe on future estimated mortality rates and life expectancies. Our paper allows to quantify this impact and provides actuaries and actuarial associations with a framework to generate scenarios of future mortality under various assessments of the pandemic data point.

\end{abstract}


\section{Introduction}\label{sec:intro}
In December 2019, the coronavirus disease (COVID-19) originated in the Chinese city Wuhan. In the months that followed, the virus spread across the world. At the time of writing, about 75 million positive cases and 1\ 403\ 245 deaths have been identified in Europe.\footnote{Numbers are retrieved from \url{https://www.statista.com/statistics/1102209/coronavirus-cases-development-europe/} and \url{https://www.statista.com/statistics/1102288/coronavirus-deaths-development-europe/} and represent the situation at October 31, 2021.} The United Kingdom has the highest absolute number of reported COVID-19 deaths in Europe (141\ 609), followed by Italy (132\ 224) and France (118\ 758). Belgium has reported 26\ 083 deaths.\footnote{These numbers of COVID-19 deaths come from the COVID-19 Dashboard by the Center for Systems Science and Engineering (CSSE) at Johns Hopkins University (JHU) on November 2, 2021, see \url{https://www.arcgis.com/apps/opsdashboard/index.html\#/bda7594740fd40299423467b48e9ecf6}.} The announcements and roll-out of the four COVID-19 vaccines approved by the European Medicine Agency (i.e.~from BioNTech and Pfizer, Moderna, AstraZeneca and Johnson \& Johnson\footnote{See \url{https://ec.europa.eu/info/live-work-travel-eu/coronavirus-response/safe-covid-19-vaccines-europeans_en} for an overview of the approved, European COVID-19 vaccines and those currently under development, as well as corresponding references.}) have led to a sharp decline in the number of COVID-19 deaths in Europe. We aim to outline the impact of the COVID-19 pandemic on a stochastic multi-population mortality projection model, such as IA$\mid$BE 2020 published by the Institute of Actuaries in Belgium \citep{IABE2020} and AG2020 by the Royal Dutch Actuarial Association \citep{KAG2020}. Further, we assess the impact of the pandemic on scenarios generated for future mortality rates with such multi-population mortality models. 

The COVID-19 pandemic has impacted mortality in multiple ways. The disease itself has led to an increase in the number of deaths, especially at the higher ages. However, measures taken by the governments worldwide also impacted mortality in a positive way, leading to less traffic or work-related accidents in 2020 and an increased awareness of sanitary precautions leading to a mild flu season in the winter of 2020-2021. While this paper puts focus on the mortality projection standard for the Belgian population, as documented in IA$\mid$BE 2020, we acknowledge some other, recent contributions that aim at assessing the impact of COVID-19 on mortality forecasts. \citet{KAG2020} performs a sensitivity analysis that shows the impact of the pandemic on the Dutch cohort life expectancies in $2021$ by feeding virtual deaths and exposures in 2019-2020 to the AG2020 model. \citet{milliman} use \citet{KAG2018} as a starting point and investigate the impact of four different COVID-19 scenarios on the Dutch best estimate mortality table published in 2018. They create these COVID-19 scenarios by multiplying the mortality rates in 2020 (and later) with a shock factor. These shocks are defined for a particular age bucket as (a fraction of) the ratio of the observed death rate in the first 23 weeks of 2020 to the average of the observed death rates in the first 23 weeks in earlier years. Next, the Continuous Mortality Investigations (CMI) in the UK provide regular updates on the excess of deaths and mortality in the United Kingdom due to COVID-19.\footnote{See \url{https://www.actuaries.org.uk/learn-and-develop/continuous-mortality-investigation/cmi-working-papers/mortality-projections}.} In their updates (CMI working papers 137, 143, 147) they adjust the calibration process of the CMI\_2020 mortality model to enable the weighting of observations. In the core version of CMI\_2020, a weight of zero is attached to to the 2020 data point and a weight of $100\%$ to all other years.  

As outlined in \citet{IABE2020}, IA$\mid$BE 2020 calibrates a mortality model of type Li \& Lee \citep{LiandLee} on the data set with the annual observed number of deaths, $d_{x,t}$, and the corresponding exposures to risk, $E_{x,t}$, registered at individual ages. More specifically, IA$\mid$BE 2020 puts focus on a set of countries over the calibration period $1988$-$2018$ (European trend) and 1988-2019 (Belgian trend) with age range 0-90. While data collected in age buckets are swiftly available from the Short-Term Mortality Fluctuations ([STMF]) Data series or Eurostat, the publication of individual age statistics takes more time. Therefore we propose a protocol to move from weekly mortality data registered in age buckets to annual mortality data at individual ages. This is a first contribution of our paper to the existing literature on mortality modelling. \citet{rizzi2015efficient} use the composite link model to ungroup coarsely grouped data, but their underlying smoothness assumption would lead to a smooth exposure and death curve. However, our protocol attempts to capture the age-specific pattern within these curves, based on historically observed data. We use our protocol to create (virtual) exposures and deaths at individual ages for the year 2020 (or earlier if necessary), leading to an extended multi-population mortality data set for the years 1988-2020 on an individual age basis. In this paper, we then assess the impact of COVID-19 on the calibration of and projections with a stochastic multi-population mortality model using this extended data set. Related work is in~\citet{schnurch2021impact}, who investigate the impact of COVID-19 on the parameters, forecasts and implied present values of life contingent liabilities with the simple Lee-Carter mortality model \citep{LeeCarter} using mortality data collected in age buckets. Our work extends the current literature by focusing on a multi-population instead of a single population mortality model, calibrated on data collected at individual ages. Moreover, we investigate the COVID-19 impact on future mortality rates and life expectancies by proposing ways to weigh the impact of this pandemic data point in either the calibration or projection set-up.

This paper is organised as follows. First, Section~\ref{sec:data} introduces some basic concepts and discusses notation. Moreover, we list the data sources that provide us with weekly and annual death counts and exposures at the level of individual ages or age buckets. In Section~\ref{sec:technical.description} we then introduce the model specifications, the assumed time dynamics and the calibration and projection methodology of the stochastic multi-population mortality projection model used in \citet{IABE2020} and \citet{KAG2020}. In addition, we specify the multi-population data set and the calibration period in the mortality model by Li \& Lee to model the Belgian mortality rates. In Section~\ref{sec:virtdata1920} we create the COVID-19 impacted data set of deaths and exposures until the year $2020$ by ungrouping the data collected in age buckets to data at the level of individual ages. Next, we recalibrate the multi-population mortality model underlying the IA$\mid$BE 2020 framework and present different methods to deal with the 2020 pandemic data point in Section~\ref{sec:methods2020}. We also assess the impact of COVID-19 on the cohort life expectancy in 2020. We conclude in Section~\ref{sec:outlook}. Technical details are deferred to the Appendix. We list the data sources in Appendix~\ref{sec:overview}. Appendix~\ref{sec:create.exp} describes the construction of the virtual exposure points $E_{x,t}$ for ages 0-90 and years 2019-2020. In Appendix~\ref{sec:create.deathcounts}, we construct the death counts $d_{x,t}$ for the same set of ages and years.

\section{Data and notation}\label{sec:data} 
\paragraph{Basic concepts.} Let $q_{x,t}$ denote the mortality rate at exact age $x$ in year $t$. This mortality rate $q_{x,t}$ refers to the probability that an $x$ year old person who was born on January 1 of year $t - x$ and is still alive at January 1 of year $t$, dies within the next year. In addition, let $\mu_{x,t}$ denote the force of mortality, i.e.~the instantaneous rate of mortality at exact age $x$ in year $t$. We assume that the force of mortality is constant in between exact ages and years, i.e.~$\mu_{x+s,t+s} = \mu_{x,t}$ for $s \in [0,1)$. Under this piecewise constant force of mortality assumption we obtain
$$ q_{x,t} = 1 - \exp\left(-\mu_{x,t}\right).$$
Stochastic mortality models, as mentioned in Section~\ref{sec:intro}, often model a transformation of the force of mortality $\mu_{x,t}$ or the mortality rate $q_{x,t}$.

\paragraph{Data sources: annually, at individual ages.} Li \& Lee's stochastic multi-population mortality projection model \citep{LiandLee}, as considered in this paper, models the logarithm of $\mu_{x,t}$ using mortality data on a collection of European countries. Hereto, mortality data are collected over a certain calibration period $\mathcal{T}$ and a range of ages $\mathcal{X}$. We use annual mortality data consisting of the observed number of deaths $d_{x,t}$ and the observed exposures to risk $E_{x,t}$, as available from sources like the Human Mortality Database ([HMD])\footnote{This database is our primary database and can be consulted at \url{https://www.mortality.org/}.}, Eurostat\footnote{Eurostat is the statistical office of the European Union, see \url{https://ec.europa.eu/eurostat}.} or an official national statistics institute like Statbel in Belgium.\footnote{Statbel is the Belgian statistical office, see \url{https://statbel.fgov.be/en}.} The latter data source is typically used to extract the most recent mortality information from the country of interest, in our case Belgium. 

\paragraph{Data sources: weekly, in age buckets.} The data sources discussed above typically report annual mortality statistics at the individual age level with a significant delay (for some countries with a delay of several years). To evaluate the impact of a pandemic shock on a mortality projection model, we therefore need other data sources that report mortality statistics in a more timely manner. Hereto, we consult the Short-Term Mortality Fluctuations ([STMF]) Data series \footnote{This information can be explored using the visualization toolkit on \url{https://mpidr.shinyapps.io/stmortality/}.} and Eurostat.\footnote{Eurostat provides weekly death statistics at \url{https://ec.europa.eu/eurostat/web/COVID-19/data}.} With only a minor delay of a few weeks, they provide weekly mortality data registered in age buckets $[x_i,x_{j}]$ rather than at the individual age level. To be compliant with the design of a stochastic multi-population mortality model, Section~\ref{sec:virtdata1920} outlines a protocol to transform these weekly mortality statistics in age buckets into annual death counts and exposures at individual ages. We use the following notations (for now, we leave out gender $g$ in our notation):
\begin{align*}
\bucket{d}, \hspace{1cm} \bucket{E} \hspace{0.5cm} \text{and} \hspace{0.5cm} \bucket{m},
\end{align*}
for the death counts, exposures and (central) death rates in age bucket $[x_i,x_{j}]$ in week $w$ in year $t$ respectively. Here, the week $w \in \{1,2,3, \ldots, 52, (53)\}$.\footnote{The years 1992, 1998, 2004, 2009, 2015 and 2020 contain 53 weeks instead of the usual 52 weeks (ISO 8601 standard).} We now further explain the weekly mortality information retrieved from the STMF data series and Eurostat:
\begin{itemize}
\item[] \textbf{STMF.} The STMF data series reports death counts $\bucket{d}$ and death rates $\bucket{m}$ in age buckets. The weekly death rates $\bucket{m}$ are derived from the weekly death counts $\bucket{d}$ and exposures $\bucket{E}$ using the following relationship:
\begin{align}\label{eq:deathrate}
\bucket{m} = \dfrac{\bucket{d}}{\bucket{E}}.
\end{align}
The STMF data series reports the weekly mortality statistics in large age buckets:
$$[0,14],\hspace{0.3cm} [15,64],\hspace{0.3cm} [65,74],\hspace{0.3cm} [75,84],\hspace{0.3cm} 85+.$$
The exposures $\bucket{E}$, used to calculate the death rates $\bucket{m}$ in Equation~\eqref{eq:deathrate}, are based on the observed annual exposures $E_{x,t}$ registered at individual ages, as reported by the HMD. However, for the most recent years, the exposures are not available yet and estimates have to be made. The STMF data series \mbox{documentation\footnote{This documentation can be consulted on \url{https://www.mortality.org/Public/STMF_DOC/STMFNote.pdf}.}} explains the construction of these unknown exposures at the level of individual ages. In addition, the STMF data series assumes a constant weekly exposure per year, per age bucket and per gender. The weekly exposure $\bucket{E}$, as reported in the STMF data series in age buckets, is the yearly (estimated) exposure divided by 52 and aggregated over the individual ages in the age bucket $[x_i,x_j]$. 
\vspace{0.1cm}
\item[] \textbf{Eurostat.} Next to the HMD and its STMF data series project, Eurostat lists valuable data sets related to death counts, useful to assess the impact of COVID-19 on mortality rates. Eurostat does not report any information about weekly exposures. From Eurostat we obtain the death counts $\bucket{d}$ by week, gender and 5-year age bucket.\footnote{See \url{https://appsso.eurostat.ec.europa.eu/nui/show.do?data set=demo_r_mwk_05&lang=en}.} The $19$ respective age buckets are
$$[0,4],\hspace{0.3cm} [5,9],\hspace{0.3cm} [10,14],\hspace{0.3cm} [15,19],\hspace{0.3cm} \ldots,\hspace{0.3cm} [85,89],\hspace{0.3cm} 90+.$$
For many countries, the STMF reported death counts correspond to the aggregated death counts reported by Eurostat. If this correspondence holds true, the data from Eurostat is more preferable due to the smaller age buckets, which eventually leads to a more accurate transition towards death counts at individual ages, necessary in the stochastic multi-population mortality projection model. For data quality reasons we only use the Eurostat reported weekly death counts in the small age buckets whenever their aggregated death counts correspond to the ones reported in the STMF data series. This is the case for all countries, except Germany\footnote{Eurostat only provides weekly death counts for Germany for age buckets of length 10.}, France and the United Kingdom.  
\end{itemize}

\begin{figure}[ht!]
\centering
\includegraphics[width = 0.8\textwidth]{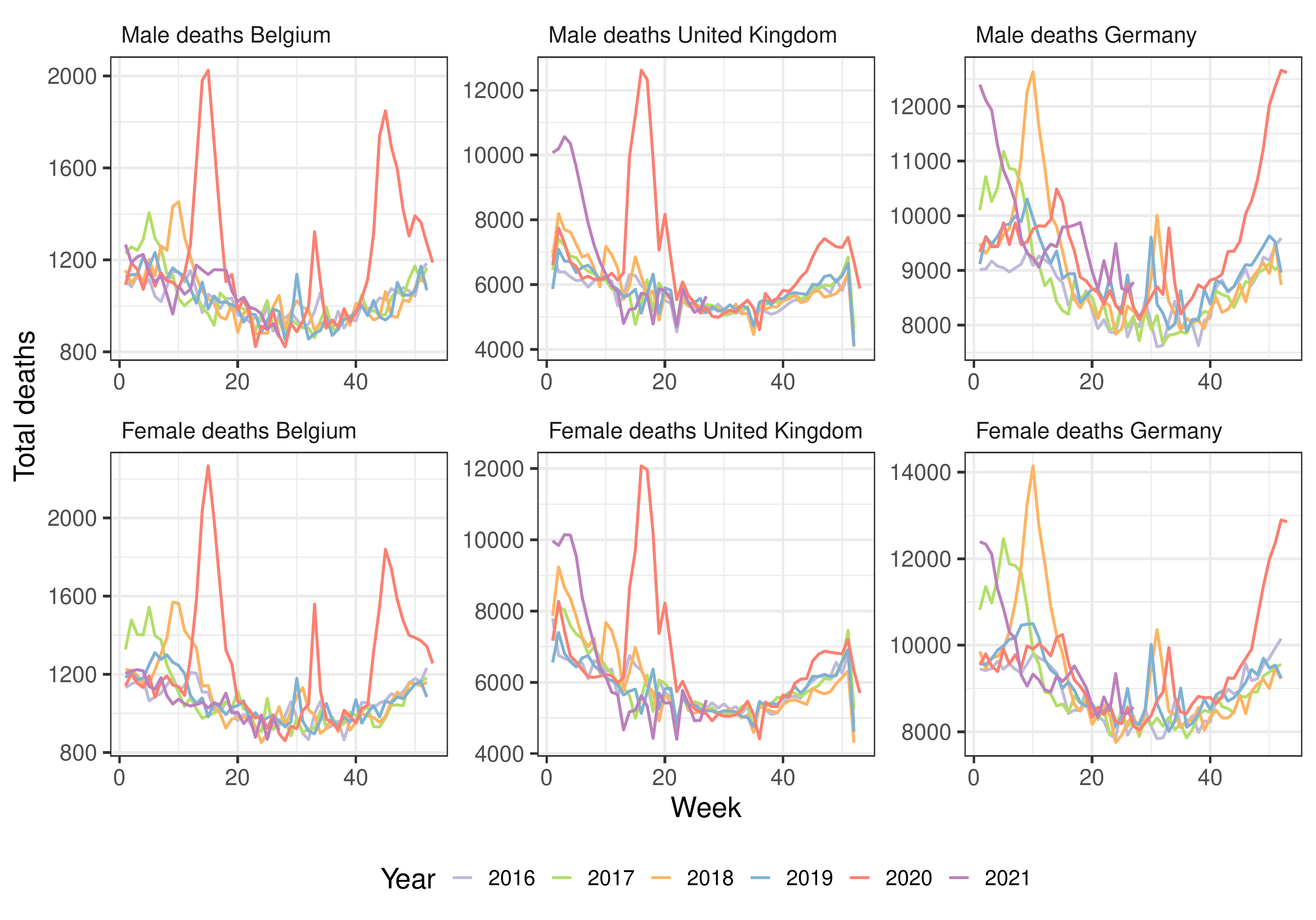}
\caption{Total weekly deaths in Belgium (left), United Kingdom (middle) and Germany (right) in the years 2016-2020 and 2021 (first 27 weeks) for males (top) and females (bottom). Eurostat (Belgium) and STMF (United Kingdom and Germany) data.\label{fig:eurodeathsbel}}
\end{figure}

Figure~\ref{fig:eurodeathsbel} illustrates the number of deaths per week for the years 2016-2021 for Belgium, United Kingdom and Germany. We clearly observe (multiple) peaks corresponding to various COVID-19 waves.  

\section{A stochastic multi-population mortality standard of type \mbox{Li \& Lee}}\label{sec:technical.description}
\citet{antonio2017producing} provide an in-depth discussion motivating the use of the Li \& Lee model as a mortality projection standard for the Dutch and Belgian population. This motivation is threefold. First, a stochastic projection model is preferred to be able to quantify the uncertainty in mortality and life expectancy forecasts and to generate scenarios of future mortality. Second, by combining country-specific data with data from other, similar European countries, the multi-population approach is more robust compared to the single population strategy. Third, the choice for the Li \& Lee model is based on an extensive, comparative analysis of the various mortality models discussed in \citet{CairnsNAAJ}, \citet{HabRenshIME2011}, \citet{borger2014modeling}, \citet{VanBerkum2014} and \citet{haberman2014longevity}. Models in this comparative analysis have been compared in terms of statistical criteria (in- and out-of-sample) and biological reasonableness. The goal of this paper is to evaluate the impact of a pandemic shock on mortality forecasts from this model of type Li \& Lee. The tools and methods to achieve this, as we develop in Section~\ref{sec:methods2020}, can be generalized to other types of mortality models.

\subsection{The Li \& Lee mortality model}
\paragraph{Specification.} The Li \& Lee mortality model \citep{LiandLee} structures the logarithm of the force of mortality for a country of interest $c$ as:
\begin{equation} \label{eq:belmuxtformula}
\begin{aligned}
\ln \mu_{x,t}^{\text{c}} &= \ln \mu_{x,t}^{\text{T}}+\ln \tilde{\mu}_{x,t}^{\text{c}} \\
\ln \mu_{x,t}^{\text{T}} &= A_x + B_xK_t  \\
\ln \tilde{\mu}_{x,t}^{\text{c}} &= \alpha_x + \beta_x \kappa_t.
\end{aligned}
\end{equation}
We recognize two Lee \& Carter specifications \citep{LeeCarter}, one to model a common mortality trend (driven by $\mu_{x,t}^{\text{T}}$) and one to model the country-specific deviation from this common trend (driven by $\tilde{\mu}_{x,t}^{\text{c}}$). This common trend reflects the global mortality trend over a collection of countries.

\paragraph{Calibration.} We calibrate this mortality model on annual data with the observed number of deaths, $d_{x,t}$, and the corresponding exposures to risk, $E_{x,t}$, over a specified age range $\mathcal{X} = \{0,1,\ldots,90\}$ and over a specified calibration period $\mathcal{T} = \{t_{\text{min}},\ldots, t_{\max}\}$. We hereby impose a Poisson distribution on the number of deaths random variable $D_{x,t}$ \mbox{\citep{BrouhnsDenuit}} and apply a conditional maximum likelihood approach \citep{Li2013}.

\begin{itemize}
\item[1.] In a first step, we calibrate the parameters $A_x,\: B_x$ and $K_t$ in the common mortality trend by assuming that the total number of deaths random variable $D_{x,t}^T$ follows a Poisson distribution with mean $\mu_{x,t}^T \cdot E_{x,t}^T$. Hereto, we maximize the following Poisson log-likelihood, conditional on the estimates obtained in step 1.:
\begin{align*}
\max_{A_x,B_x,K_t} \sum_{x \in \mathcal{X}} \sum_{t \in \mathcal{T}} \left( d_{x,t}^{\text{T}} \log\left(\mu_{x,t}^{\text{T}}\right) - E_{x,t}^{\text{T}}\mu_{x,t}^{\text{T}} \right),
\end{align*}
where $d_{x,t}^{\text{T}}$ and $E_{x,t}^T$ are the observed number of deaths and exposures respectively, aggregated over the collection of countries. Further, $\mu_{x,t}^T = \exp(A_x + B_x K_t)$. We impose some constraints on the Lee-Carter parameters to avoid identification problems:
$$ \displaystyle \sum_{x \in \mathcal{X}} B_x^2 = 1, \hspace{0.5cm} \sum_{t \in \mathcal{T}} K_t = 0.$$
\item[2.] In a second step, we calibrate the country-specific parameters $\alpha_x,\: \beta_x$ and $\kappa_t$ by assuming that the number of deaths random variable $D_{x,t}^c$, in the country of interest $c$, follows a Poisson distribution with mean $\mu_{x,t}^c \cdot E_{x,t}^c$. Hereto, we maximize the Poisson log-likelihood:
\begin{align*}
\max_{\alpha_x,\beta_x,\kappa_t} \sum_{x \in \mathcal{X}} \sum_{t \in \mathcal{T}} \left( d_{x,t}^{\text{c}} \log\left(\mu_{x,t}^{\text{c}}\right) - E_{x,t}^{\text{c}}\mu_{x,t}^{\text{c}} \right),
\end{align*}
where $d_{x,t}^{\text{c}}$ and $E_{x,t}^c$ are the observed number of deaths and exposures in country $c$ respectively. Further, we have $\mu_{x,t}^c = \mu_{x,t}^{\text{T}} \cdot \exp(\alpha_x + \beta_x \kappa_t)$. In line with step 1, we again impose some identifiability constraints on the country-specific Lee-Carter parameters:
$$ \displaystyle \sum_{x \in \mathcal{X}} \beta_x^2 = 1, \hspace{0.5cm} \sum_{t \in \mathcal{T}} \kappa_t = 0.$$
\end{itemize}

\subsection{The time dynamics} \label{subsec:timedyna}
\paragraph{Specification.} The time dynamics of the common period effect, $K_t$, are modelled with a Random Walk with Drift ([RWD]). The country-specific period effect, $\kappa_t$, follows an AR(1) process with intercept. These choices are based on the work of \citet{IABE2020} and \citet{KAG2020}. Hence, we use the following bivariate time series models:
\begin{equation} \label{eq:timedyn}
\begin{aligned}
K_{t} &= K_{t-1} + \theta + \epsilon_{t} \\
\kappa_{t} &= c + \phi \kappa_{t-1} + \delta_{t},
\end{aligned}
\end{equation}
for males ($M$) and females ($F$) separately, with $t \in \{t_{\text{min}}+1,\ldots, t_{\text{max}}\}$. We assume that the four-dimensional vectors of noise terms $(\epsilon_{t}^M, \delta_t^M, \epsilon_{t}^F, \delta_t^F)$ are independent over time and follow a four-dimensional Gaussian distribution with mean $(0,0,0,0)$ and covariance matrix $\boldsymbol C$. We denote:
\begin{align*}
\resizebox{\hsize}{!}{$\boldsymbol{Y}_t  = \begin{pmatrix}
K_{t}^M - K_{t-1}^M \\ \kappa_{t}^M \\ K_{t}^F - K_{t-1}^F \\ \kappa_{t}^F 
\end{pmatrix} \in \mathbb{R}^{4\times 1}, \hspace{0.25cm}
\boldsymbol{X}_t = \begin{pmatrix}
1 & 0 & 0          & 0 & 0 & 0   \\
0 & 1 & \kappa_{t-1}^M & 0 & 0 & 0   \\
0 & 0 & 0          & 1 & 0 & 0   \\
0 & 0 & 0          & 0 & 1 & \kappa_{t-1}^F   
\end{pmatrix} \in \mathbb{R}^{4\times 6}, \hspace{0.25cm}
\boldsymbol{\Psi} = \begin{pmatrix}
\theta^M \\ c^M \\ \phi^M \\ \theta^F \\ c^F \\ \phi^F
\end{pmatrix}  \in \mathbb{R}^{6\times 1}$,}
\end{align*}
where $t \in \{t_{\min} +1,\ldots,  t_{\max}\}$. Using this notation, $\boldsymbol{Y}_t - \boldsymbol{X}_t \boldsymbol{\Psi}$ represents the four-dimensional vector of noise terms $(\epsilon_t^M, \delta_t^M, \epsilon_t^F, \delta_t^F)^t$ at time $t$.

\paragraph{Calibration.} Inspired by \citet{KAG2020}, we estimate the time series parameters $\boldsymbol{\Psi}$ and the covariance matrix $\boldsymbol{C}$ on the calibrated $\hat{K}_t$ and $\hat{\kappa}_t$ parameters, jointly for males and females, by maximizing the four-dimensional Gaussian log-likelihood:\footnote{We use the \texttt{nlminb}-function in the \texttt{stats}-package of \texttt{R}.}
\begin{equation} \label{eq:4variatell}
\begin{aligned}
l(\boldsymbol{\Psi}, \boldsymbol{C}) &= \log \left( \displaystyle \prod_{t=t_{\min}+1}^{t_{\max}} \dfrac{1}{\sqrt{(2\pi)^4 |\boldsymbol{C}|}} e^{-\frac{1}{2} (\boldsymbol{Y}_t - \boldsymbol{X}_t \boldsymbol{\Psi})^t \boldsymbol{C}^{-1} (\boldsymbol{Y}_t - \boldsymbol{X}_t \boldsymbol{\Psi})}\right) \\
&=  -\left(\left|\mathcal{T}\right|-1\right) \left(2\log 2\pi + 0.5\log |\boldsymbol{C}|\right) - \frac{1}{2}\displaystyle \sum_{t=t_{\min}+1}^{t_{\max}}(\boldsymbol{Y}_t - \boldsymbol{X}_t \boldsymbol{\Psi})^t \boldsymbol{C}^{-1} (\boldsymbol{Y}_t - \boldsymbol{X}_t \boldsymbol{\Psi}) \\
&= -\left(\left|\mathcal{T}\right|-1\right) \left(2\log 2\pi + 0.5\log |\boldsymbol{C}|\right) - \frac{1}{2}\displaystyle \sum_{t=t_{\min} + 1}^{t_{\max}} tr\left[\boldsymbol{C}^{-1}(\boldsymbol{Y}_t - \boldsymbol{X}_t \boldsymbol{\Psi})(\boldsymbol{Y}_t - \boldsymbol{X}_t \boldsymbol{\Psi})^t \right],
\end{aligned}
\end{equation}
where $\left|\mathcal{T}\right|$ is the number of years in the calibration period $\mathcal{T}$ and where $tr(\cdot)$ is the trace function applied to a matrix. We denote the calibrated time series parameters as $\widehat{\boldsymbol{\Psi}}$ and the calibrated covariance matrix as $\widehat{\boldsymbol{C}}$.

\subsection{Generating future paths of mortality rates and life expectancies}\label{subsec:futurepaths}
\paragraph{Mortality rates.} We now use these calibrated time dynamics to generate future paths for the country-specific mortality rates $\hat \mu_{x,t}^{\text{c}}$. We consider a projection period $t \in \{t_{\max} +1,...,T\}$. Hereto, we start from the calibrated period effects in the last year of the calibration period $\mathcal{T}$, i.e.~($\hat{K}_{t_{\max}}^M$, $\hat{\kappa}_{t_{\max}}^M$, $\hat{K}_{t_{\max}}^F$, $\hat{\kappa}_{t_{\max}}^F$). We then take random draws $(\epsilon_{t,i}^M, \delta_{t,i}^M, \epsilon_{t,i}^F, \delta_{t,i}^F)$ for $i \in \{1,...,n\}$ and $t \in \{t_{\max} +1,...,T\}$ from the fitted Gaussian distribution with mean $(0,0,0,0)$ and covariance matrix $\widehat{\boldsymbol{C}}$. In Algorithm~\ref{algorithm:simulation}, we obtain future paths of the calibrated period effects for males and females using Equation \eqref{eq:timedyn}. Note that $\hat{K}^g_{t_{\max},i} = \hat{K}^g_{t_{\max}}$ and $\hat{\kappa}^g_{t_{\max},i} = \hat{\kappa}^g_{t_{\max}}$ for all $i$ and each gender $g$.

\begin{algorithm}[ht!]
	    \KwIn{$\big(\hat{K}_{t_{\max}}^M, \hat{\kappa}_{t_{\max}}^M, \hat{K}_{t_{\max}}^F, \hat{\kappa}_{t_{\max}}^F\big)$, $\widehat{\boldsymbol{\Psi}}$, $\widehat{\boldsymbol{C}}$}
    \KwOut{$n$ future paths for the period effects}
    \SetAlgoLined
        \For{$i$ \KwTo{} $1,\ldots,n$ }{
        	\For{$t$ \KwTo{} $t_{\max}+1, \ldots, T$ }{
        	$\big(\epsilon^M_{t,i},\delta^M_{t,i}, \epsilon^F_{t,i},\delta^F_{t,i}\big)$  := mvrnorm$\big($n = 4, mu = (0,0,0,0), Sigma = $\widehat{\boldsymbol{C}}\big)$ \\
            	\For{$g$ \KwTo{} $M,F$ }{
$\hat{K}^g_{t,i} = \hat{K}^g_{t-1,i} + \hat{\theta}^g + \epsilon^g_{t,i}$ \\
$\kappa^g_{t,i} = \hat{c}^g + \hat{\phi}^g \hat{\kappa}^g_{t-1,i} + \delta^g_{t,i}$.
        		}
        	}    
        }
    \caption{Generating future paths for the calibrated period effects $(\hat{K}_t^M, \hat{\kappa}_t^M, \hat{K}_t^F, \hat{\kappa}_t^F)$.}
    \label{algorithm:simulation}
\end{algorithm}

Using Equation~\eqref{eq:belmuxtformula} and the calibrated Li \& Lee parameters $\hat{A}_x, \:\hat{B}_x,\:\hat{K}_t, \: \hat{\alpha}_x,\: \hat{\beta}_x$ and $\hat{\kappa}_t$, we can generate future paths for the country-specific mortality rates. Let us denote $\hat{q}_{x,t,i}^{\text{c}}$ and $\hat{\mu}_{x,t,i}^{\text{c}}$ for the $i$-th generated value of the country-specific mortality rate $\hat{q}_{x,t}^{\text{c}}$ and the country-specific force of mortality $\hat{\mu}_{x,t}^{\text{c}}$ respectively. Then, we obtain:
\begin{align}\label{eq:linkmuq}
\hat q_{x,t,i}^{\text{c}} = 1 - \exp\left(-\hat \mu_{x,t,i}^{\text{c}}\right),
\end{align} 
with $x \in \mathcal{X}$, $t \in \{t_{\max} +1,...,T\}$ and $i \in \{1,...,n\}$. Having obtained a scenario for the mortality rates for ages 0-90 in a future year, we close the generated mortality rates until age 120 using the method of \citet{Kannisto}. We refer to \citet{IABE2020} for a detailed explanation of this method.

\paragraph{Life expectancy.}  We obtain future paths of the period and cohort life expectancies of an $x$ year old in year $t$ \citep{pitacco} as:
\begin{equation} \label{eq:LifeExp}
\begin{aligned}
\hat{e}_{x,t,i}^{\text{per}} &= \frac{1-\exp{(-\hat{\mu}_{x,t,i})}}{\hat{\mu}_{x,t,i}}+\sum_{k\geq 1} \left(\prod_{j=0}^{k-1} \exp{(-\hat{\mu}_{x+j,t,i})}\right)\frac{1-\exp{(-\hat{\mu}_{x+k,t,i})}}{\hat{\mu}_{x+k,t,i}}, \\
\hat{e}_{x,t,i}^{\text{coh}} &= \frac{1-\exp{(-\hat{\mu}_{x,t,i})}}{\hat{\mu}_{x,t,i}}+\sum_{k\geq 1} \left(\prod_{j=0}^{k-1} \exp{(-\hat{\mu}_{x+j,t+j,i})}\right)\frac{1-\exp{(-\hat{\mu}_{x+k,t+k,i})}}{\hat{\mu}_{x+k,t+k,i}}.
\end{aligned}
\end{equation}

\subsection{The Li \& Lee mortality model for the Belgian population}\label{subsec:mortprojbe}
IA$\mid$BE 2020 is based on a mortality model of type Li \& Lee and puts focus on Belgium as the country of interest ($c = \text{BEL}$). The common trend in Equation~\eqref{eq:belmuxtformula} is a European mortality trend calibrated on a set of countries with a Gross Domestic Product per capita above the European average in 2018.\footnote{See \url{https://data.worldbank.org/indicator/NY.GDP.PCAP.CD}.} As such, the multi-population data set combines mortality data from Belgium, The Netherlands, Luxembourg, Norway, Switzerland, Austria, Ireland, Sweden, Denmark, Germany, Finland, Iceland, United Kingdom and France. Further, IA$\mid$BE 2020 calibrates the mortality model on annual data registered at the level of individual ages, from the HMD, Eurostat and the Belgian statistical institute Statbel. IA$\mid$BE 2020 calibrates the parameters in the European mortality trend $\mu_{x,t}^T$ on a range of years 1988-2018 and the Belgium-specific mortality trend $\mu_{x,t}^{\text{BEL}}$ on the range of years 1988-2019.

The aim of this paper is to recalibrate the Li \& Lee mortality model on a multi-population data set consisting of the same set of countries, but on a calibration period from 1988 to 2020. Hereto, we use the STMF data series and Eurostat, providing weekly mortality statistics collected in age buckets. Neither the STMF data series nor Eurostat report granular mortality information on Ireland. While Ireland is part of the set of countries defined in IA$\mid$BE 2020 to calibrate the common European mortality trend, we exclude this country in the COVID-19 impact assessment covered in this paper. However, given the exposure size of the Irish population, we do not expect that this has a major impact on the results obtained with the multi-population mortality model.

\section{Transforming weekly mortality data in age buckets to annual mortality data at individual ages}\label{sec:virtdata1920}
Our strategy to evaluate the impact of COVID-19 on a stochastic multi-population mortality projection model of type Li \& Lee, adheres to the design principles of the model. As explained in Section~\ref{subsec:mortprojbe}, we have to supplement our multi-population data set with mortality data for the most recent years 2019-2020. Hereto, we use the weekly mortality statistics in age buckets, retrieved from the STMF data series and Eurostat. We propose a protocol to make the transition from the weekly deaths $\bucket{d}$ and exposures $\bucket{E}$ collected in age buckets to the required format. Section~\ref{subsec:weektoyear} discusses the transition from weekly to annual deaths and exposures collected in age buckets. In section~\ref{sec:buckettoindividual}, we convert the annual deaths $d_{[x_i,x_j],t}$ and exposures $E_{[x_i,x_j],t}$ registered in age buckets to data at the level of individual ages. 

\subsection{From weekly to annual mortality data registered in age buckets}\label{subsec:weektoyear}
\paragraph{STMF.} The Short-Term Mortality Fluctuations Data series, as discussed in Section~\ref{sec:data}, assume a constant weekly exposure per year, per age bucket and per gender. When going from the weekly exposures available from the STMF data series to annual exposures, we simply multiply the weekly exposures with a factor $52$, i.e.~the number of weeks in a year:
$$ E_{[x_i,x_j],t} = 52 \bucket{E},$$
where $E_{[x_i,x_j],t}$ now denotes the total annual exposure in year $t$ for age bucket $[x_i,x_j]$. 

\paragraph{STMF and Eurostat.} The weekly mortality data sources, i.e.~the STMF data series and Eurostat, follow the definition of `week' given by the ISO week date system, which is part of the ISO-8601 date and time standard.\footnote{See \url{https://www.iso.org/iso-8601-date-and-time-format.html}.} In this system, a year consists of 52 or 53 full weeks. When a year $t$ consists of 53 weeks instead of the usual 52 weeks,\footnote{The years 1992, 1998, 2004, 2009, 2015 and 2020 contain 53 weeks.} we adjust the calculation of the yearly death counts to be compliant with the HMD and Eurostat death counts registered at individual ages:
$$ d_{[x_i,x_j],t} = \frac{52}{53} \displaystyle \sum_{w=1}^{53} d_{[x_i,x_j],t,w}.$$
In addition, the STMF data series lists death counts and death rates for Northern Ireland, England and Wales and Scotland separately. A simple aggregation of their death counts leads to the death counts of the United Kingdom as a whole.

\subsection{Ungrouping data from age buckets to individual ages}\label{sec:buckettoindividual}
\begin{figure}[ht!]
\centering
\includegraphics[width = 0.8\textwidth]{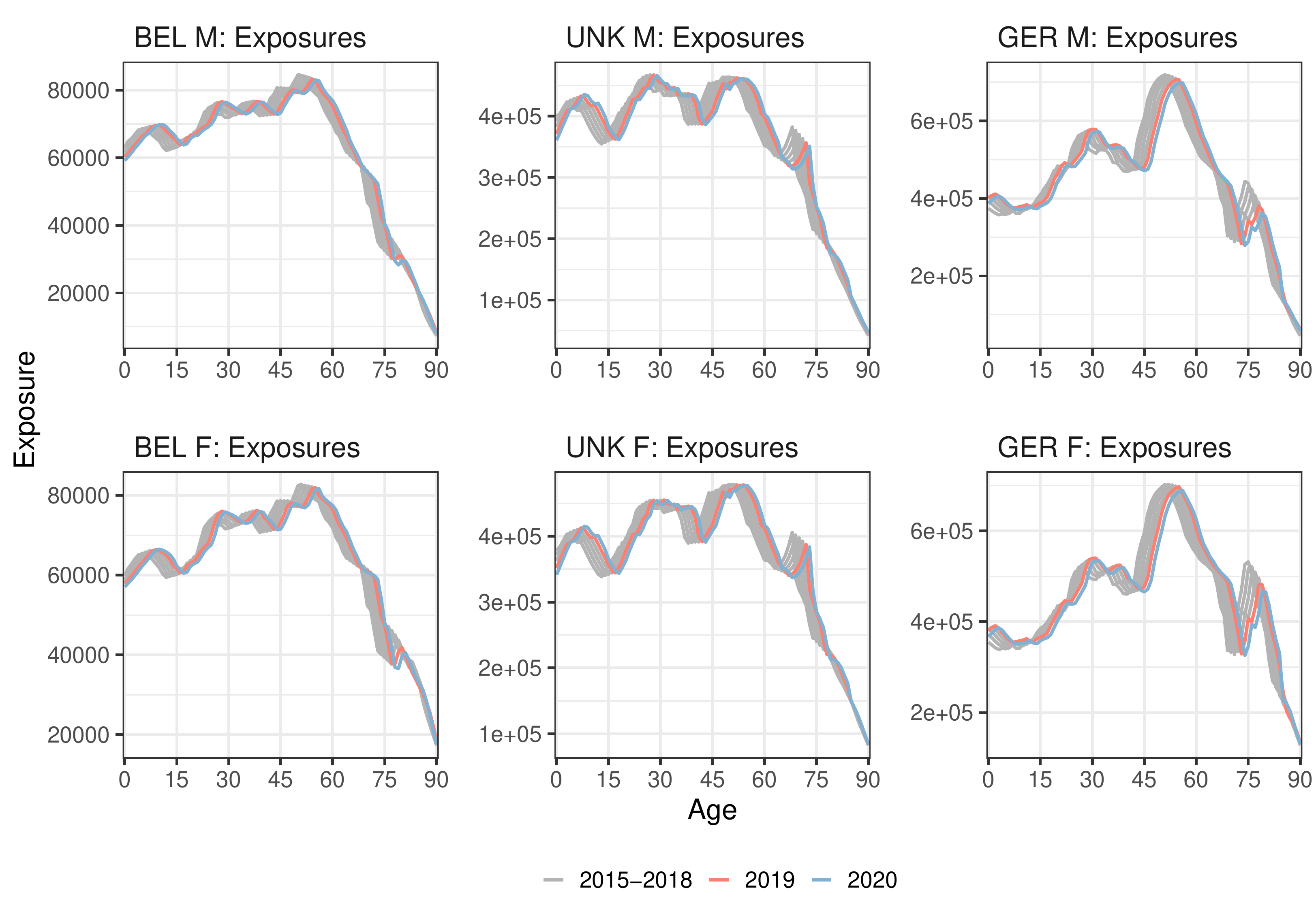}
\caption{(Virtual) annual exposures $E_{x,t}$ for males (top) and females (bottom) in Belgium (left), the United Kingdom (middle) and Germany (right) as a function of age, years $2015$-$2020$. Exposures for the years 2015-2019 (Belgium), 2015-2018 (United Kingdom) and 2015-2019 (Germany) are directly available from the HMD or Eurostat. However, for the year 2020 (Belgium and Germany) and the years 2019-2020 (United Kingdom), we transform the weekly exposures collected in age buckets from the STMF data series to exposures at individual ages.\label{fig:expbelunkger}}
\end{figure}

\begin{figure}[ht!]
\centering
\includegraphics[width = 0.8\textwidth]{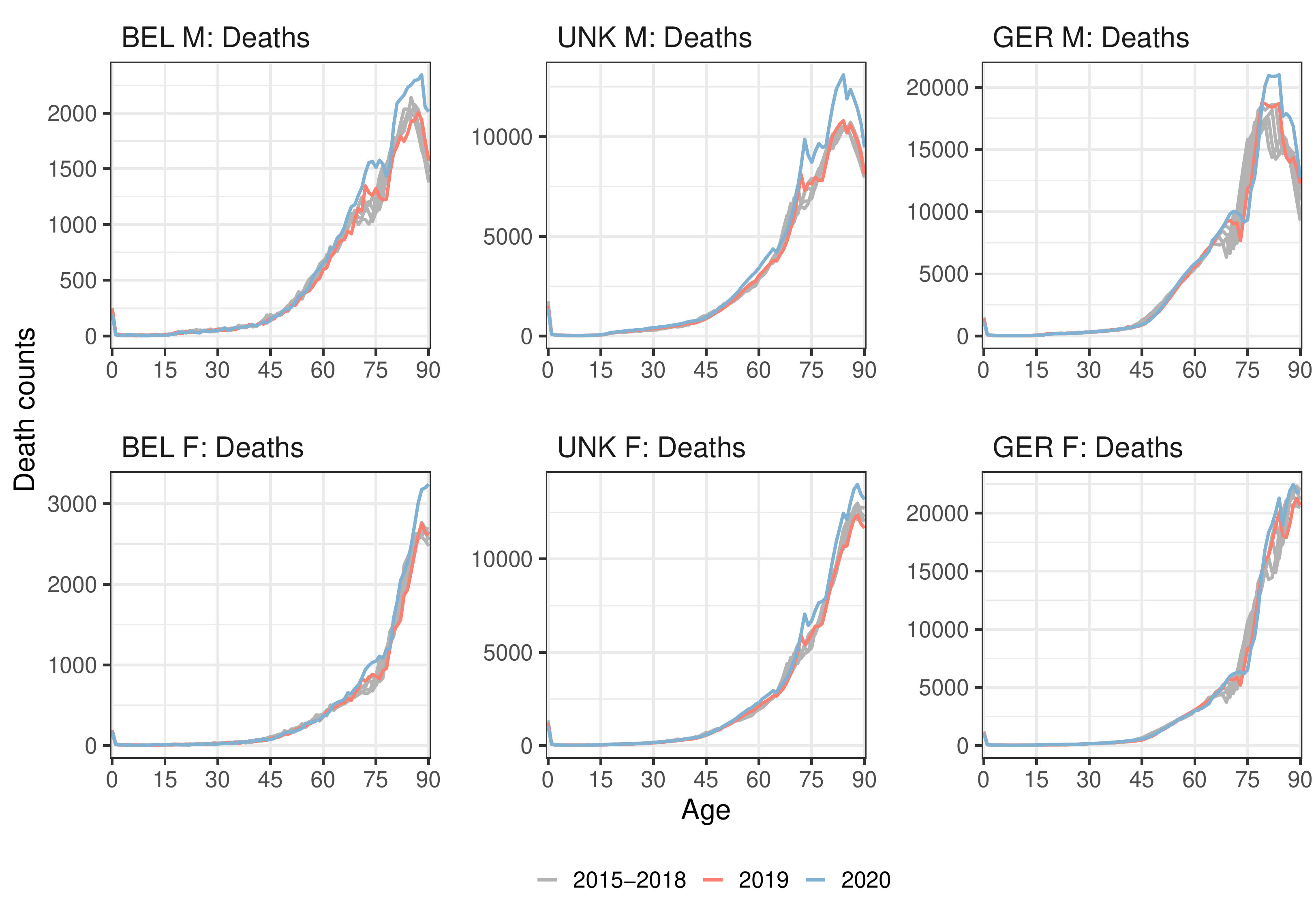}
\caption{(Virtual) annual death counts $d_{x,t}$ for males (top) and females (bottom) in Belgium (left), the United Kingdom (middle) and Germany (right) as a function of age, years $2015$-$2020$. Death counts for the years 2015-2020 (Belgium), 2015-2018 (United Kingdom) and 2015-2019 (Germany) are directly available from the HMD, Eurostat or Statbel. However, for the year 2020 (Germany) and the years 2019-2020 (United Kingdom), we transform the weekly deaths collected in age buckets from the STMF data series or Eurostat to deaths at individual ages.\label{fig:deathsbelunkger}}
\end{figure}

We start from the annual deaths $d_{[x_i,x_j],t}$ and exposures $E_{[x_i,x_j],t}$ collected in age buckets, as obtained from Section~\ref{subsec:weektoyear}. We then define a protocol to ungroup the data in age buckets to data at the level of individual ages. \citet{rizzi2015efficient} introduce a method that ungroups histograms (or coarsely grouped data), using a composite linked model with a penalty to ensure the smoothness of the underlying distribution. Their strategy is not able to capture typical patterns in the evolution of the exposures or death counts over time. As an example, Figures~\ref{fig:expbelunkger} and \ref{fig:deathsbelunkger} in this paper clearly show the evolution of certain spikes in the exposure and death curve over time, i.e.~the spikes are moving to the right by one age each subsequent year. These observed spikes or patterns within an age bucket cannot be captured by the method of \citet{rizzi2015efficient}. Therefore, we propose an alternative strategy that is capable to pick up these spikes. At the same time we ensure that the sum of the individual, ungrouped number of deaths and exposures in an age bucket $[x_i,x_j]$ corresponds to the exposures and deaths in that age bucket, reported in the STMF data series or by Eurostat. We call the newly created annual deaths and exposures at the level of individual ages, virtual deaths and exposures.

\begin{figure}[ht!]
\centering
\includegraphics[width = 0.8\textwidth]{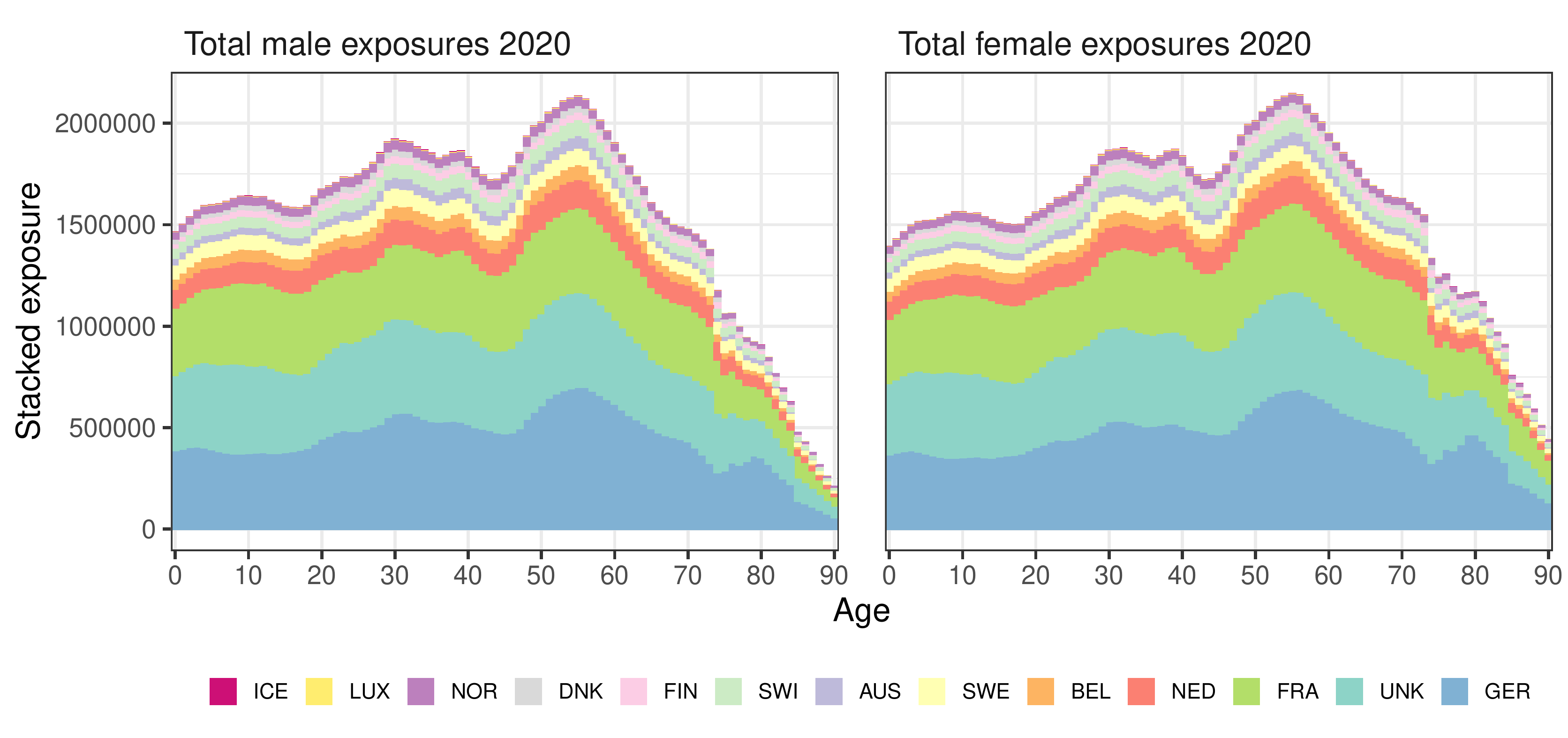}
\caption{Stacked male (left) and female (right) exposure for the combined thirteen European countries in 2020 as a function of age. Exposures for the year 2020 are directly available from the HMD for Denmark. However, for Germany, United Kingdom, France, Netherlands, Belgium, Sweden, Austria, Switzerland, Finland, Norway, Luxembourg and Iceland we transform the weekly exposures in age buckets from the STMF data series to exposures at individual ages for the year 2020. \label{fig:stackedeuexp}}
\end{figure}

\paragraph{Protocol to ungroup $E_{[x_i,x_j],t}$.} To obtain annual exposures at individual ages $E_{x,t}$ in an unknown year $t$, we take the STMF or Eurostat exposure data $E_{[x_i,x_j],t}$ for the corresponding age buckets. We then allocate these exposures $E_{[x_i,x_j],t}$ to exposures at individual ages $E_{x,t}$ by applying a piecewise scaling of the known exposure curve from a previous year. Appendix~\ref{sec:create.exp} explains our approach in full detail. Figure~\ref{fig:stackedeuexp} visualizes the stacked exposures at ages 0-90 in the year 2020 for all 13 European countries under consideration for males (left) and females (right). Figure~\ref{fig:expbelunkger} shows the evolution of the (virtual) annual exposures for Belgium, the United Kingdom and Germany. The exposures in the year 2020 (and 2019 for the United Kingdom) are created using our approach. 

\begin{figure}[ht!]
\centering
\includegraphics[width = 0.8\textwidth]{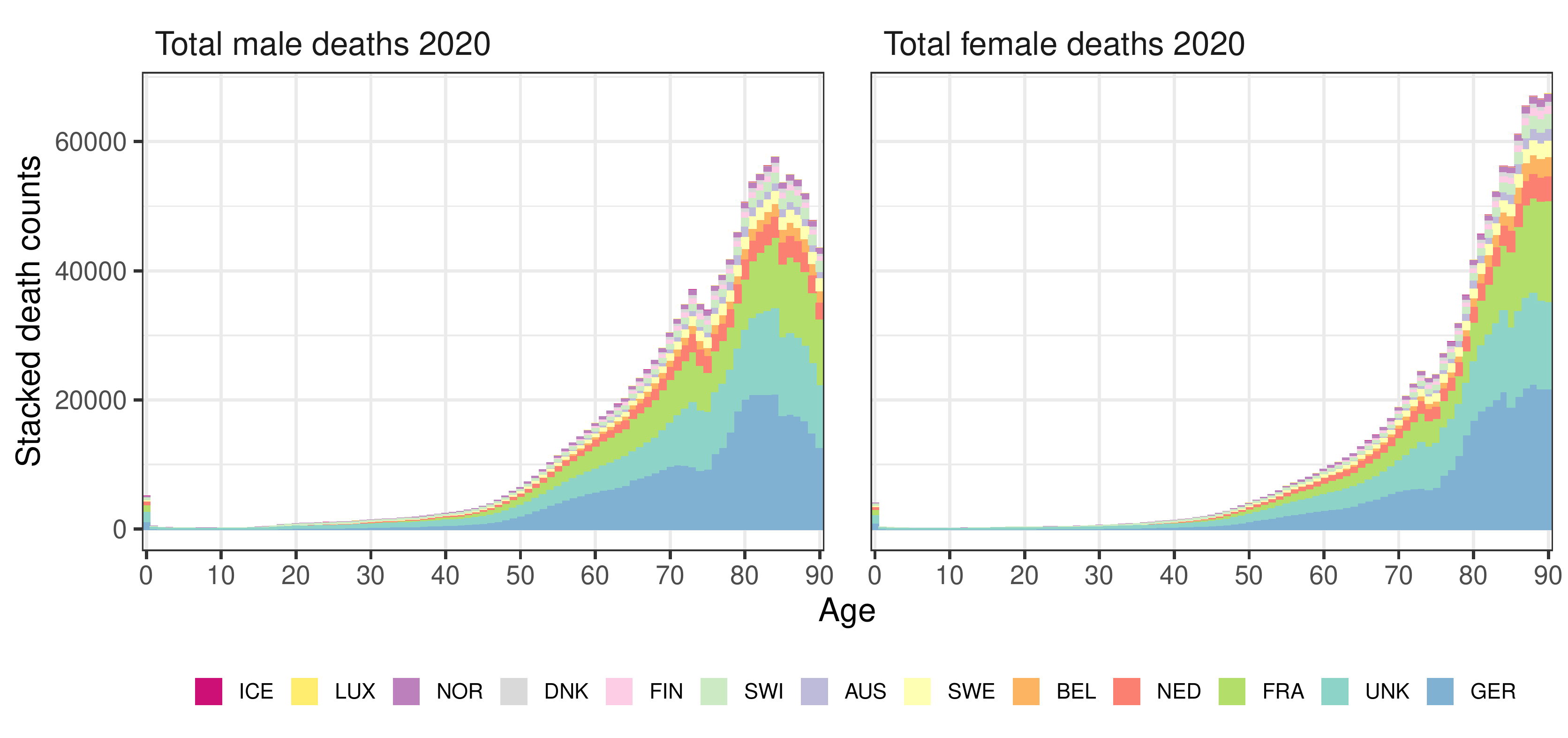}
\caption{Stacked male (left) and female (right) death counts for the combined thirteen European countries in 2020 as a function of age. Death counts for the year 2020 are directly available from the HMD and Statbel for Denmark and Belgium respectively. However, for Germany, United Kingdom, France, Netherlands, Sweden, Austria, Switzerland, Finland, Norway, Luxembourg and Iceland we transform the weekly deaths in age buckets from the STMF data series or Eurostat to deaths at individual ages for the year 2020.\label{fig:deathseustacked}}
\end{figure}

\paragraph{Protocol to ungroup $\boldsymbol{d_{[x_i,x_j],t}}$.} We create the annual death counts at individual ages $d_{x,t}$ in year $t$ for which the statistics at individual ages have not yet been published. To do this, we take the raw death counts $d_{[x_i,x_j],t}$ by age buckets reported in the STMF data series or by Eurostat and allocate these to individual ages. Hereto, we extrapolate a multi-population mortality model (e.g.~the IA$\mid$BE 2020 model) to obtain mortality rate estimates in year $t$. Then we combine these estimates with the (virtual) exposures from year $t$ to obtain virtual death counts in year $t$. Appendix~\ref{sec:create.deathcounts} again provides the technical details. Figure~\ref{fig:deathseustacked} shows the stacked European number of deaths in 2020 for males (left panel) and females (right panel). Figure~\ref{fig:deathsbelunkger} shows the evolution of deaths over time for Belgium, the United Kingdom and Germany. The excess of deaths in 2020 due to COVID-19 is clearly visible at old ages.

\subsection{Applying the protocol to the multi-population data set to recalibrate the Li \& Lee mortality model for the Belgian population}
Table~\ref{tab:overview} in Appendix~\ref{sec:overview} indicates for which countries and for which years the data set must be supplemented with information from the weekly deaths and exposures registered in age buckets from the STMF data series or Eurostat. For those years and those countries, we apply the protocol from Sections~\ref{subsec:weektoyear} and \ref{sec:buckettoindividual} to move from the weekly deaths and exposures in age buckets to annual data at individual age level. We then recalibrate the Li \& Lee model for the Belgian population on an extended data set up to and including the year 2020. The United Kingdom is the only country for which we have to create virtual exposures and death counts at individual ages for both the years 2019 and 2020. Next, we only have to create virtual exposures for the year 2020 for Belgium since we retrieve the death counts in 2020 at individual ages from Statbel.\footnote{See \url{https://statbel.fgov.be/sites/default/files/files/documents/bevolking/5.4\%20Sterfte\%2C\%20levensverwachting\%20en\%20doodsoorzaken/5.4.1\%20Sterfte/Verdeling\%20overlijdens\%20per\%20leeftijd\%20en\%20geslacht\%20sinds\%201992_NL.xlsx}} Moreover, at the time of writing, the HMD already provides deaths and exposures for Denmark in 2020 at the level of individual ages. For all other considered European countries we have to create virtual deaths and exposures for the year 2020. 

\section{Assessing the impact of a pandemic shock on the multi-population mortality model} \label{sec:methods2020}
The IA$\mid$BE 2020 mortality model \citep{IABE2020} is calibrated on data from 1988-2018 (European trend) and 1988-2019 (Belgium-specific trend). Using the protocol from Section~\ref{sec:virtdata1920}, we are able to collect data from 1988-2020 for all 13 countries under consideration in this multi-population mortality model. As a starting point we recalibrate the mortality model on the data set up to and including the pandemic year 2020. However, since the mortality shock takes place in the last year of our calibration period $\mathcal{T}$, it has a major impact on the estimated drift term in the assumed random walk with drift process for the European period effect (see Equation~\eqref{eq:timedyn}). This, in turn, severely impacts the mortality and life expectancy forecasts, as the year 2020 is the starting year to generate future scenarios of mortality (Section~\ref{subsec:futurepaths}). 

With the roll-out of the four COVID-19 vaccines, approved by the European Medicine Agency, a more optimistic scenario is that post-pandemic mortality rates will continue their improvement at a rate that is similar to pre-pandemic levels. Section~\ref{subsec:method1} introduces a method to put this scenario into practice. We limit the weight of the pandemic data point (the year 2020) in the four-dimensional Gaussian log-likelihood (see Equation~\eqref{eq:4variatell}) to estimate the time series parameters in Equation~\eqref{eq:timedyn}. However, this strategy uses the pandemic 2020 observations as starting point to generate future mortality projections. Therefore, we still observe a fairly large difference between the short-term future mortality rates and life expectancies when comparing the results obtained with the original IA$\mid$BE 2020 model and the recalibrated mortality model.

The impact of the pandemic year 2020 concentrates on old age mortality rates, as Figure~\ref{fig:deathsbelunkger} illustrates. The stochastic multi-population mortality projection model by Li \& Lee \citep{LiandLee}, as specified in Equation~\eqref{eq:belmuxtformula}, cannot capture this age-specific effect of the pandemic on the mortality rates. Indeed, while the upward jump in the estimated common European period effect $\hat{K}_t$ is driven by the deterioration in mortality rates for the older ages, it actually has the largest effect on the fitted mortality rates for the younger ages (see Figure~\ref{fig:eubelage_A1} later on). Therefore, the Li \& Lee model overestimates the observed mortality rates at the younger ages and underestimates the deterioration of the mortality rates at the older ages in the pandemic year 2020. 

In light of the above discussion, Section~\ref{subsec:method2} introduces a method that slightly changes the model specifications in Equation~\eqref{eq:belmuxtformula}. We here impose that the fitted death rates in the pandemic year 2020 are equal to a weighted average of the observed death rates in the year 2020 and the pre-pandemic rates in 2019. By giving a zero weight to the observed death rates in 2020, we can investigate the situation where we completely ignore the COVID-19 pandemic and are of the belief that the post-pandemic death rates in 2021 immediately recover to pre-pandemic levels.  

\subsection{Limiting the time series likelihood contribution of the pandemic data point} \label{subsec:method1}
\begin{figure}[h!]
\centering
\includegraphics[width = 0.78\textwidth]{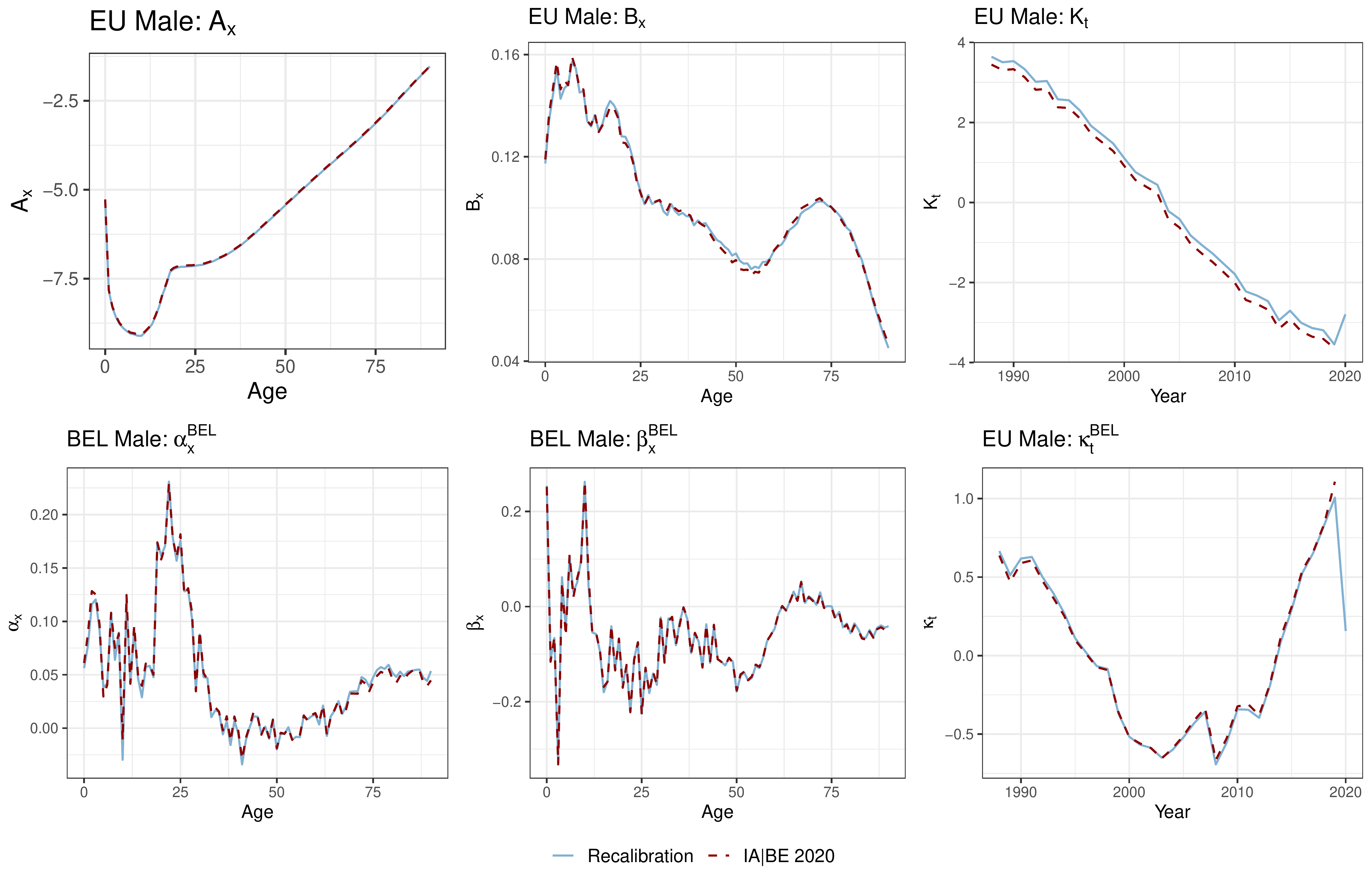}
\includegraphics[width = 0.78\textwidth]{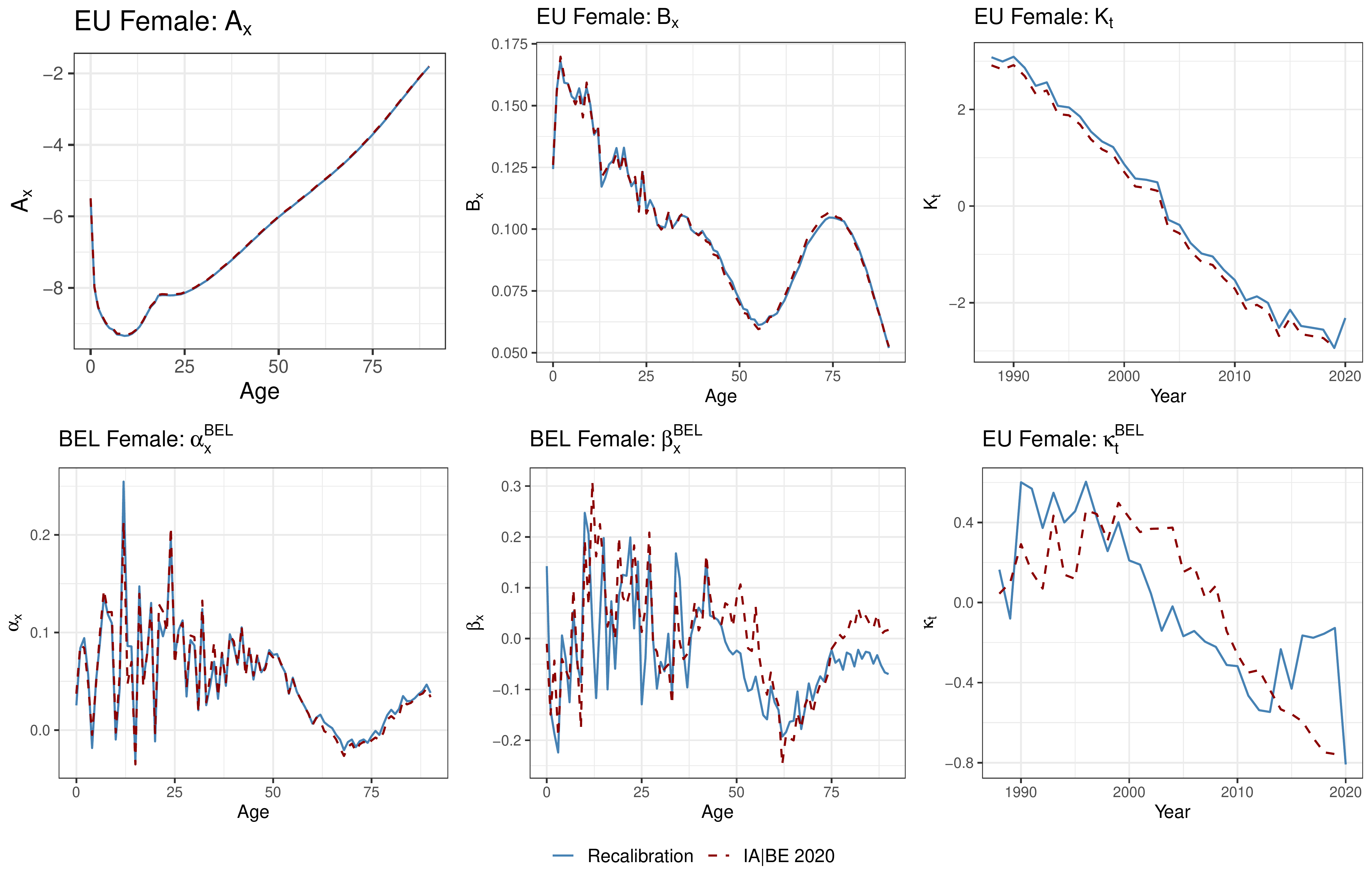}
\caption{The European and Belgium-specific Li \& Lee parameters $\hat{A}_x$, $\hat{B}_x$, $\hat{\alpha}_x$, $\hat{\beta}_x$, $\hat{K}_t$ and $\hat{\kappa}_t$ for males and females. The blue line corresponds to the Li \& Lee model calibrated on data from the years 1988-2020 and contains virtually created deaths and exposures for the years 2019-2020. The dark red, dashed line shows the calibrated Li \& Lee parameters in the original IA$\mid$BE 2020 model (calibration period 1988-2018(9)). \label{fig:eubelage_A1}}
\end{figure}

We first perform a simple recalibration of the mortality model on the data set including the (virtually created) death counts and exposure points up to the year 2020. This recalibration is completely in line with the model choices and design principles underneath IA$\mid$BE 2020. Figure~\ref{fig:eubelage_A1} shows the recalibrated Li \& Lee parameters for males (top panels) and females (bottom panels). We visualize these together with the parameter estimates from the original IA$\mid$BE 2020 model. We do not observe any substantial differences in the calibrated age-dependent parameters $\hat{A}_x,\: \hat{B}_x$ (European trend), $\hat{\alpha}_x$ and $\hat{\beta}_x$ (Belgian deviation). The recalibrated common period effect $\hat{K}_t$ reveals a clear upward jump in the year 2020 for both males and females to account for the increase in observed mortality rates (see Figures~\ref{fig:eurodeathsbel} and \ref{fig:deathsbelunkger}). The sharp decline in the male Belgium-specific period effect $\hat{\kappa}_t$ in the year 2020 implies that the male Belgian mortality rates do not diverge further from the European mortality rates. In addition, we observe larger differences between the original and the recalibrated Belgium-specific period effect for females, i.e.~$\hat{\kappa}_t^F$.

We use the same specification of the time dynamics as in Section~\ref{subsec:timedyna}. However, we now limit the contribution of the pandemic data point in the four-dimensional Gaussian log-likelihood to estimate the four-dimensional time series for $(\hat{K}_t^M, \hat{\kappa}_t^M, \hat{K}_t^F, \hat{\kappa}_t^F)$ (see Equation~\eqref{eq:4variatell} in Section~\ref{subsec:timedyna}). Hereto, we introduce weights in the Gaussian log-likelihood:
\begin{equation} \label{eq:4variatellweights}
l(\boldsymbol{\Psi}, \boldsymbol{C}) = -\frac{1}{2}\displaystyle \sum_{t=1989}^{2020} \textcolor{red}{w_{t}} \cdot \left(4 \log 2\pi + \log |\boldsymbol{C}| +  tr\left[\boldsymbol{C}^{-1}(\boldsymbol{Y}_t - \boldsymbol{X}_t \boldsymbol{\Psi})(\boldsymbol{Y}_t - \boldsymbol{X}_t \boldsymbol{\Psi})^t \right]\right).
\end{equation}
We specify $w_t = 1$ for $t < 2020$ and consider 5 possible scenarios for $w_{2020}$. That is $w_{2020} \in \{0,0.25,0.50,0.75,1\}$. We follow the same approach for the projection set-up as in Section~\ref{subsec:futurepaths}. The use of a weighted log-likelihood function allows to assess the impact of limiting the contribution of the pandemic data point on the projections of the time-specific parameters, the mortality forecasts and the life expectancy predictions.

\begin{sidewaysfigure}
\begin{subfigure}{0.48\hsize}\centering
    \includegraphics[width=\hsize]{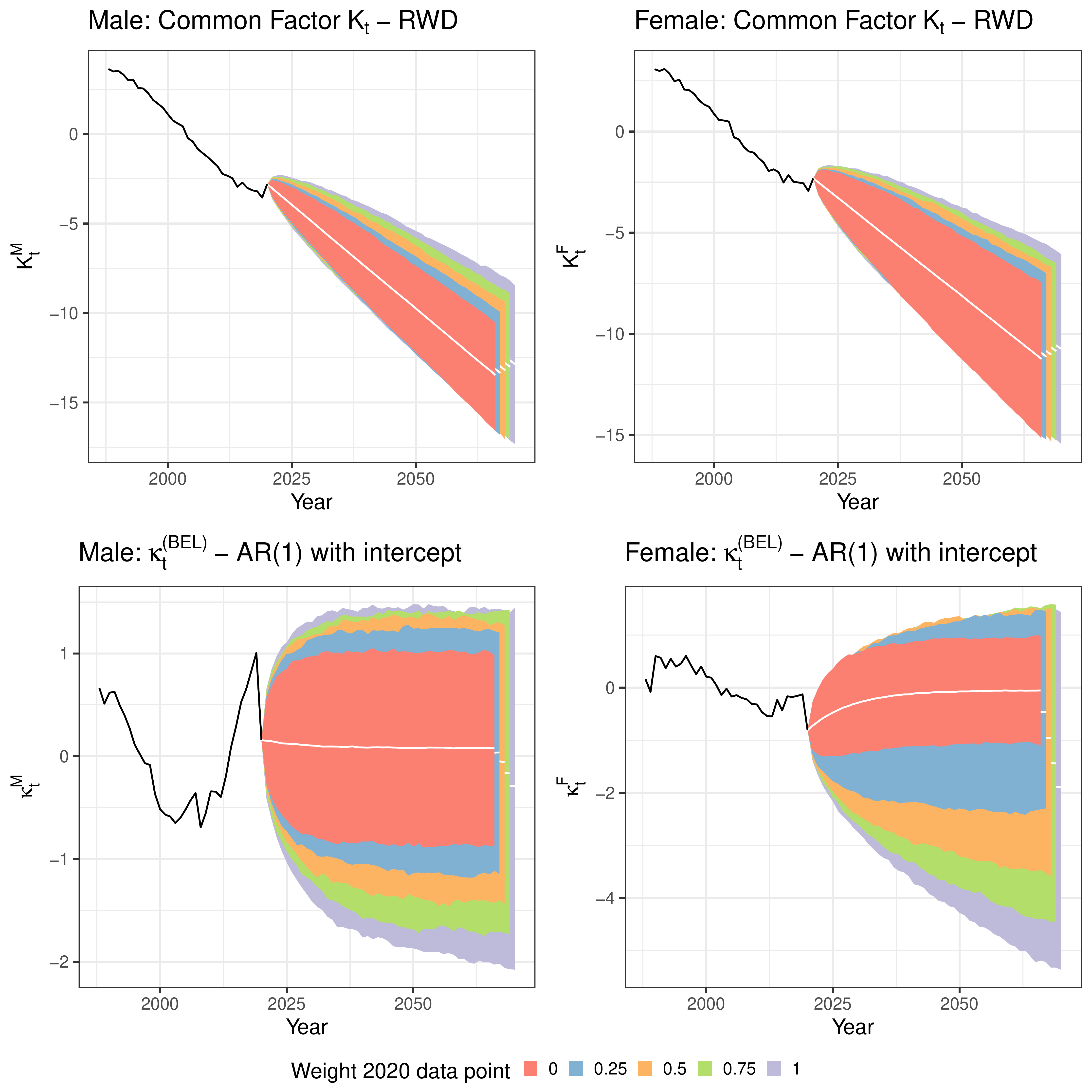}
\caption{Projection of the calibrated time dependent parameters in the Li \& Lee model: $\hat{K}_t$ (top) and $\hat{\kappa}_t$ (bottom). Male (left) and female (right) data, calibration period $1988$-$2020$, projection period $2021$-$2070$, method of Section~\ref{subsec:method1}. We show the $0.5\%$, median (white lines) and $99.5\%$ quantile, based on $10\ 000$ simulations across each weighting scenario. The black lines visualize the calibrated period effects: $\hat{K}_t$ and $\hat{\kappa}_t$.{\color{white} white space white space white space white space white space white space white space white space white space white space white space.} \label{fig:euKtmf_A1}}
\end{subfigure}%
\hfill 
\begin{subfigure}{0.48\hsize}\centering
    \includegraphics[width=\hsize]{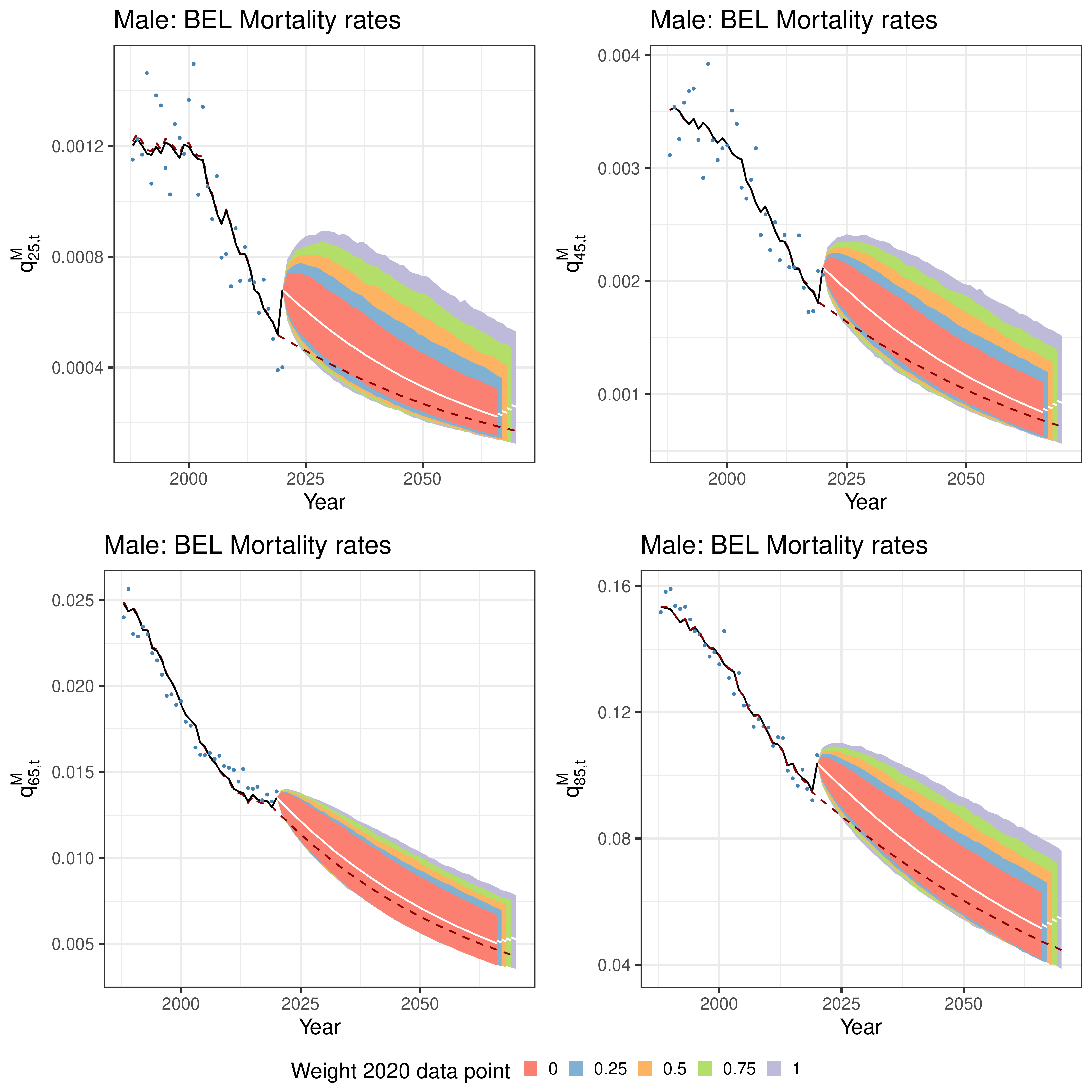}
\caption{Estimated and projected Belgian mortality rates $\hat{q}_{x,t}$. Male data, ages 25, 45 (top) and 65, 85 (bottom), calibration period 1988-2020, projection period 2021-2070, method of Section~\ref{subsec:method1}. We show the $0.5\%$, median (white lines) and $99.5\%$ quantile, based on $10\ 000$ simulations across each weighting scenario. The blue dots and the black lines represent the observed and fitted mortality rates respectively. The dark red, dashed line shows the calibrated mortality rates and the median quantile of the simulations in the original IA$\mid$BE 2020 model. \label{fig:qxtbelM25456585_A1}}
\end{subfigure}
\vspace{0.1cm}
\caption{Estimated and projected time dependent parameters (left) and Belgian mortality rates (right).}
\end{sidewaysfigure}

Figure~\ref{fig:euKtmf_A1} shows the projection of the calibrated European and Belgium-specific period effects $\hat{K}_t$ and $\hat{\kappa}_t$ for males and females. First, a lower weight for the likelihood contribution of the 2020 data point leads to less variability in the simulations. Second, lowering the 2020 weight leads to a larger (absolute) value of the drift parameter in the RWD process for the common period effect $\hat{K}_t$. In addition, it leads to a faster long term mean reversion for the Belgium-specific period effect $\hat{\kappa}_t$. This is confirmed by Table~\ref{tab:modspec_A1} in which we show the time series parameter estimates across the five different weighting scenarios of the 2020 data point. In the table, we also report the time series parameter estimates obtained in the original IA$\mid$BE 2020 model. The latter estimates should be broadly in line with the estimates of the zero weight scenario. However, some deviations may occur because the original IA$\mid$BE 2020 model uses a calibration period from 1988 to 2018 to model the European mortality trend, while our recalibration uses a calibration period from 1988 to 2020. The AR(1) parameter estimates reveal that the Belgian deviation for females becomes less stable when the 2020 data point fully contributes to the Gaussian likelihood, i.e.~the estimate $\hat{\phi}^F$ is very close to one. A potential drawback of this method to deal with the pandemic data point is that the projections jump off from the impacted (shocked) estimates $\hat{K}_{2020}$ and $\hat{\kappa}_{2020}$ in the year 2020. 

\begin{table}[ht]
\centering
\begin{tabular}{lccccccc}
\toprule   
Weight 2020 &  $\theta^M$ & $\theta^F$ & $c^M$ & $c^F$ & $\phi^M$ & $\phi^F$  \\
\midrule
0    & -0.2319 & -0.1942 & 0.0073  & -0.0060 & 0.9032 & 0.8873  \\ 
0.25 & -0.2240 & -0.1877 & 0.0021  & -0.0138 & 0.9303 & 0.9648 \\
0.50 & -0.2163 & -0.1812 & -0.0039 & -0.0196 & 0.9413 & 0.9817 \\
0.75 & -0.2087 & -0.1749 & -0.0100 & -0.0250 & 0.9471 & 0.9893 \\
1    & -0.2011 & -0.1687 & -0.0162 & -0.0302 & 0.9506 & 0.9937 \\ \\
IA$\mid$BE 2020 & -0.2285 & -0.1882 & 0.0140 & -0.0240 & 0.9682 & 0.9226 \\
\bottomrule
\end{tabular}
\caption{Time series parameter estimates, male and female data, ages $0$-$90$, years $1988$-$2020$, Method~\ref{subsec:method1}.\label{tab:modspec_A1}}
\end{table}

Figure~\ref{fig:qxtbelM25456585_A1} shows the observed, calibrated and simulated Belgian mortality rates for ages 25, 45, 65 and 85 from the recalibrated mortality model and the original (pre COVID-19) IA$\mid$BE 2020 model. We observe a rather poor in-sample fit of the male mortality rate for age $25$ in 2020 since the Li \& Lee mortality model is not able to capture the observed differences in excess of mortality between the younger and older ages. Because of the estimated, in sample increase in mortality rates at all ages, the projected mortality is clearly at a higher level in the recalibrated model compared to the projected mortality in the original IA$\mid$BE 2020 model. In addition, limiting the time series likelihood contribution of the 2020 data point, leads to lower projected mortality rates on average. The results from Figure~\ref{fig:euKtmf_A1} confirm this.

\begin{figure}[h!]
\centering
\includegraphics[width = 0.80\textwidth]{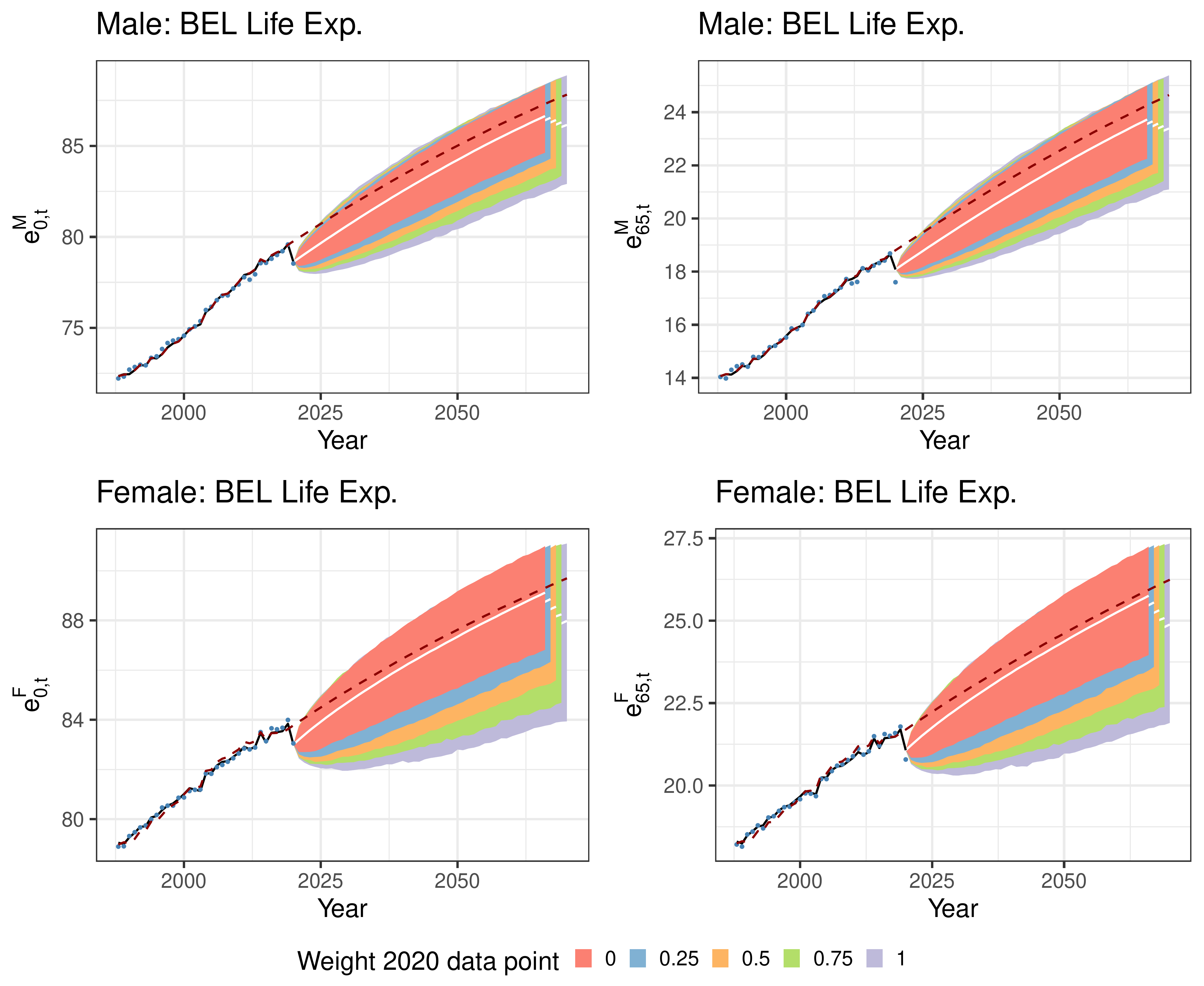}
\caption{Estimated and projected period life expectancies $\hat{e}_{x,t}$ for Belgium. Male (top) and female (bottom) data, ages 0 (left) and 65 (right), calibration period 1988-2020, projection period 2021-2070, method of Section~\ref{subsec:method1}. We show the $0.5\%$, median (white lines) and $99.5\%$ quantile (fan charts), based on $10\ 000$ simulations, across each weighting scenario. The dark red, dashed line shows the $50\%$ quantile originating from the original IA$\mid$BE 2020 model. The blue dots and black line represent the observed and fitted period life expectancies respectively.\label{fig:extbelMF065_A1}}
\end{figure}

We are now ready to assess the impact of COVID-19 on the estimated and projected period and cohort life expectancies. Figure~\ref{fig:extbelMF065_A1} depicts the projected period life expectancies of a male and female Belgian newborn (left) and a 65 year old (right) resulting from the recalibrated mortality model (fan charts) and from the original IA$\mid$BE 2020 model in \citet{IABE2020} (dark red, dashed line). Table~\ref{tab:A1CLE065} then shows the best-estimates and the $0.5\%$, median and $99.5\%$ quantiles of the 10\ 000 simulations for the cohort life expectancy of a 0 and 65 year old in 2020. Both period and cohort life expectancies are negatively impacted by COVID-19. Moreover, the long-term impact of COVID-19 reduces for the recalibrated mortality model that assigns a smaller weight to the 2020 data point in the calibration step of the time dynamics. However, we still observe a clear and pronounced short-term impact of COVID-19 on the period and cohort life expectancies across all the different weighting scenarios. In addition, we also observe less uncertainty in the life expectancy simulations when a smaller weight is allocated to the 2020 data point, as the more narrow fan charts in Figure~\ref{fig:extbelMF065_A1} indicate.

\begin{table}[ht]
\centering
\adjustbox{max width=0.9\textwidth}{%
\centering
\begin{tabular}{@{\extracolsep{4pt}}lccccc}
\toprule   
\multicolumn{2}{l}{\multirow{2}{4cm}{\textbf{Cohort life expectancy in 2020}}}  & \multicolumn{2}{c}{Males}  & \multicolumn{2}{c}{Females}\\
 \cmidrule{3-4} 
 \cmidrule{5-6} 
 & & 0 & 65 & 0 &65 \\
\midrule
\multirow{2}{*}{\shortstack[l]{Recalibration \\ \scriptsize{2020 weight = 0}}} 	& Best. Est.    & 89.28 & 19.69 & 91.42 & 22.77 \\
 &$[q_{0.5};q_{50};q_{99.5}]$ & $[87.45;89.28;90.82]$ & $[18.96;19.69;20.42]$ & $[89.30;91.41;93.16]$ & $[21.81;22.78;23.71]$ \\ \\
 \multirow{2}{*}{\shortstack[l]{Recalibration \\ \scriptsize{2020 weight = 0.25}}} 	& Best. Est.  & 89.00 & 19.63 & 91.05 & 22.57 \\
 &$[q_{0.5};q_{50};q_{99.5}]$ & $[86.86;89.00;90.77]$ & $[18.76;19.62;20.48]$ & $[88.63;91.05;93.01]$ & $[21.47;22.57;23.68]$\\ \\
\multirow{2}{*}{\shortstack[l]{Recalibration \\ \scriptsize{2020 weight = 0.50}}} 	& Best. Est.  & 88.69 & 19.56 & 90.58 & 22.44 \\
 &$[q_{0.5};q_{50};q_{99.5}]$ & $[86.26;88.68;90.69]$ & $[18.56;19.56;20.51]$ & $[87.61;90.57;92.9]$ & $[21.16;22.43;23.65]$\\ \\
\multirow{2}{*}{\shortstack[l]{Recalibration \\ \scriptsize{2020 weight = 0.75}}} 	& Best. Est.  & 88.37 & 19.50 & 90.06 & 22.32 \\
 &$[q_{0.5};q_{50};q_{99.5}]$ & $[85.56;88.35;90.57]$ & $[18.44;19.49;20.54]$ & $[86.45;90.03;92.81]$ & $[20.99;22.32;23.66]$\\ \\
\multirow{2}{*}{\shortstack[l]{Recalibration \\ \scriptsize{2020 weight = 1}}} 	& Best. Est.    & 88.05 & 19.43 & 89.52 & 22.22 \\
 &$[q_{0.5};q_{50};q_{99.5}]$ & $[84.91;88.05;90.50]$ & $[18.25;19.43;20.59]$ & $[85.57;89.51;92.61]$ & $[20.79;22.22;23.66]$\\ \\
\multirow{2}{*}{IA$\mid$BE 2020} & Best. Est.   & 89.91 & 20.38 & 91.54 & 23.14\\
 &$[q_{0.5};q_{50};q_{99.5}]$ & $[88.11;89.89;91.46]$ & $[19.57;20.37;21.17]$ & $[89.46;91.53;93.25]$ & $[22.15;23.14;24.07]$\\ \\
\bottomrule
\end{tabular}}
\caption{The cohort life expectancy for a 0 and 65 year old in 2020. The best estimate and the $0.5\%$, median and $99.5\%$ quantile obtained from 10\ 000 simulations are shown, for males and females.\label{tab:A1CLE065}}
\end{table} 

A closer look at Figure~\ref{fig:extbelMF065_A1} reveals an over-estimation of the period life expectancy in 2020 for a 65-year old. This is primarily due to the fact that the Li \& Lee model tries to achieve a good in-sample fit for both the young and the old ages. As a result the model produces higher mortality rates than observed for the young ages and lower mortality results than observed for the older ages, as confirmed by Figure~\ref{fig:qxtbelM25456585_A1} and the paragraph about mortality rates.  

\subsection{Mitigating the impact of the pandemic data point with a Lee \& Miller inspired mortality model} \label{subsec:method2}
The method discussed in Section~\ref{subsec:method1} to deal with the pandemic data point has two potential drawbacks. First, we get a poor in-sample fit of the observed mortality rates and life expectancies in the pandemic year 2020. Second, even with a low to zero weight assigned to the 2020 period effects in the time series likelihood, we obtain a clear short-term impact of COVID-19 on the predicted mortality rates and life expectancies. This scenario may be considered unrealistic in light of the effectiveness of the approved vaccines. In view of the aforementioned shortcomings, we therefore propose two modifications to the Li \& Lee mortality model.

\paragraph{Modification 1.} A first modification consists in slightly changing the model specifications of the Li \& Lee model (see Equation~\eqref{eq:belmuxtformula}). Hereto, note that the (central) death rate $m_{x,t}$ equals: 
$$ m_{x,t} = \dfrac{d_{x,t}}{E_{x,t}}.$$ 
Under the piecewise constant force of mortality assumption, introduced in Section~\ref{sec:data}, the maximum likelihood estimate of the force of mortality, $\hat{\mu}^{{\tiny \text{MLE}}}_{x,t}$, equals the observed death rate $m_{x,t}$. \citet{leemiller} adjust the Lee \& Carter model specification in such a way that the fitted forces of mortality in the last year of the calibration period $\mathcal{T}$ are equal to the observed death rates in that year. This provides a solution for the poor in-sample fit at young ages in the pandemic year 2020. We extend this idea to the Li \& Lee mortality model. We hereby reduce the degrees of freedom in the mortality model of Equation \eqref{eq:belmuxtformula} by fixing the parameter values for $A_x$ and $\alpha_x$ such that the fitted and observed country-specific mortality rates in the year 2020 match.

\paragraph{Modification 2.} Second, we mitigate the (short-term) impact of the pandemic data point on the mortality projections. We do this by relaxing the fact that the observed death rates are exactly the same as the fitted forces of mortality in the pandemic year 2020. Instead, we opt for a weighted average between the observed death rates in the last two years of the calibration period. 

\paragraph{The adjusted Lee \& Miller model specifications.} 
These two discussed modifications result in the following model specifications:\footnote{The case $\alpha_{2020} = 1$ corresponds to the Lee \& Miller mortality model as discussed in \citet{leemiller}.}
\begin{equation} \label{eq:belmuxtMiller2}
\begin{aligned}
\ln \mu_{x,t}^{\text{c}} &= \ln \mu_{x,t}^{\text{T}}+\ln \tilde{\mu}_{x,t}^{\text{c}} \\
\ln \mu_{x,t}^{\text{T}} &= \alpha_{2020} \cdot \log m_{x,2020}^{\text{T}} + \left(1-\alpha_{2020}\right) \cdot \log m_{x,2019}^{\text{T}} + B_x(K_t - K_{2020})  \\
\ln \tilde{\mu}_{x,t}^{\text{c}} &= \alpha_{2020} \cdot \log \tilde{m}_{x,2020}^{\text{c}} +  \left(1-\alpha_{2020}\right) \cdot \log \tilde{m}_{x,2019}^{\text{c}} + \beta_x (\kappa_t - \kappa_{2020}),
\end{aligned}
\end{equation}
with $t \in \mathcal{T} = \{1988, ...,2020\}$ and $x \in \mathcal{X} = \{0,1,...,90\}$. Further, for $t \in \{2019,2020\}$, $m_{x,t}^{\text{T}}$ is the observed common central death rate and $\tilde{m}_{x,t}^{\text{c}}$ the observed country-specific `central death rate' with adjusted exposure $E_{x,t}^{\text{c}} \cdot m_{x,t}^{\text{T}}$:
\begin{align}\label{eq:obsmortmiller}
m_{x,t}^{\text{T}} = \dfrac{d_{x,t}^{\text{T}}}{E_{x,t}^{\text{T}}}, \hspace{1cm} \tilde{m}_{x,t}^{\text{c}} = \dfrac{d_{x,t}^{\text{c}}}{E_{x,t}^{\text{c}} \cdot m_{x,t}^{\text{T}}}.
\end{align} 
In addition, $\alpha_{2020} \in \{0,0.25,0.50,0.75,1\}$ is the weight we assign to the observed central death rates in 2020. With the definitions in Equation~\eqref{eq:obsmortmiller}, the case $\alpha_{2020} = 1$ corresponds to the situation in which the observed country-specific death rates are equal to the fitted country-specific forces of mortality in the year 2020. As a consequence, the fitted and observed country-specific mortality rates also coincide in the year 2020. For this weighting scenario, we therefore get a perfect fit of the observed period life expectancy in the year 2020. Taking $\alpha_{2020} = 0$ corresponds to the situation where the fitted mortality rates in 2020 equal the observed mortality rates in 2019. In the latter scenario we completely ignore COVID-19 and assume that the mortality rates have not been changed over the years 2019-2020. We refer to the model, specified in Equation~\eqref{eq:belmuxtMiller2}, as the \textit{adjusted Lee \& Miller mortality model}.

\begin{figure}[h!]
\centering
\includegraphics[width = 0.8\textwidth]{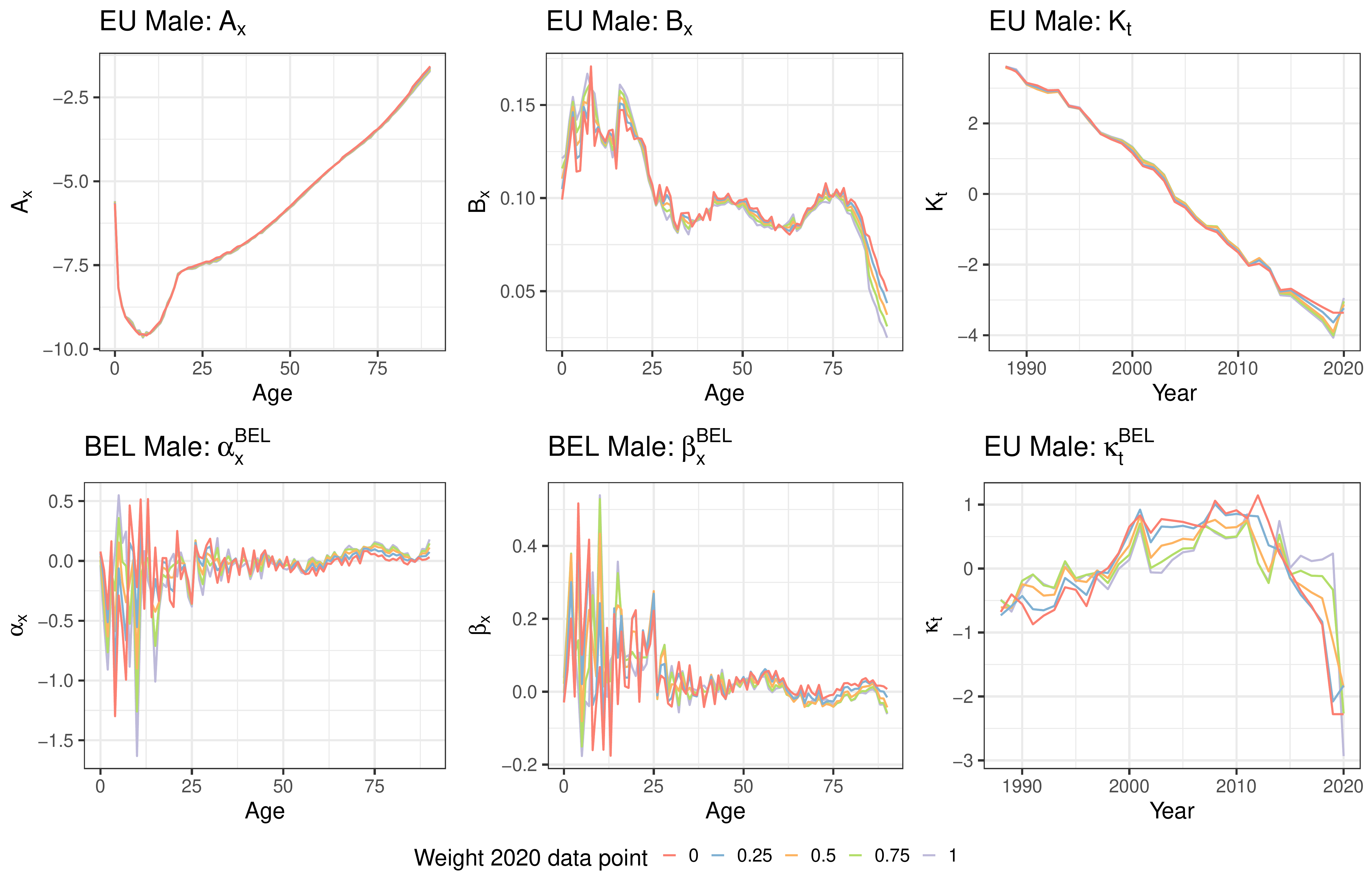}
\includegraphics[width = 0.8\textwidth]{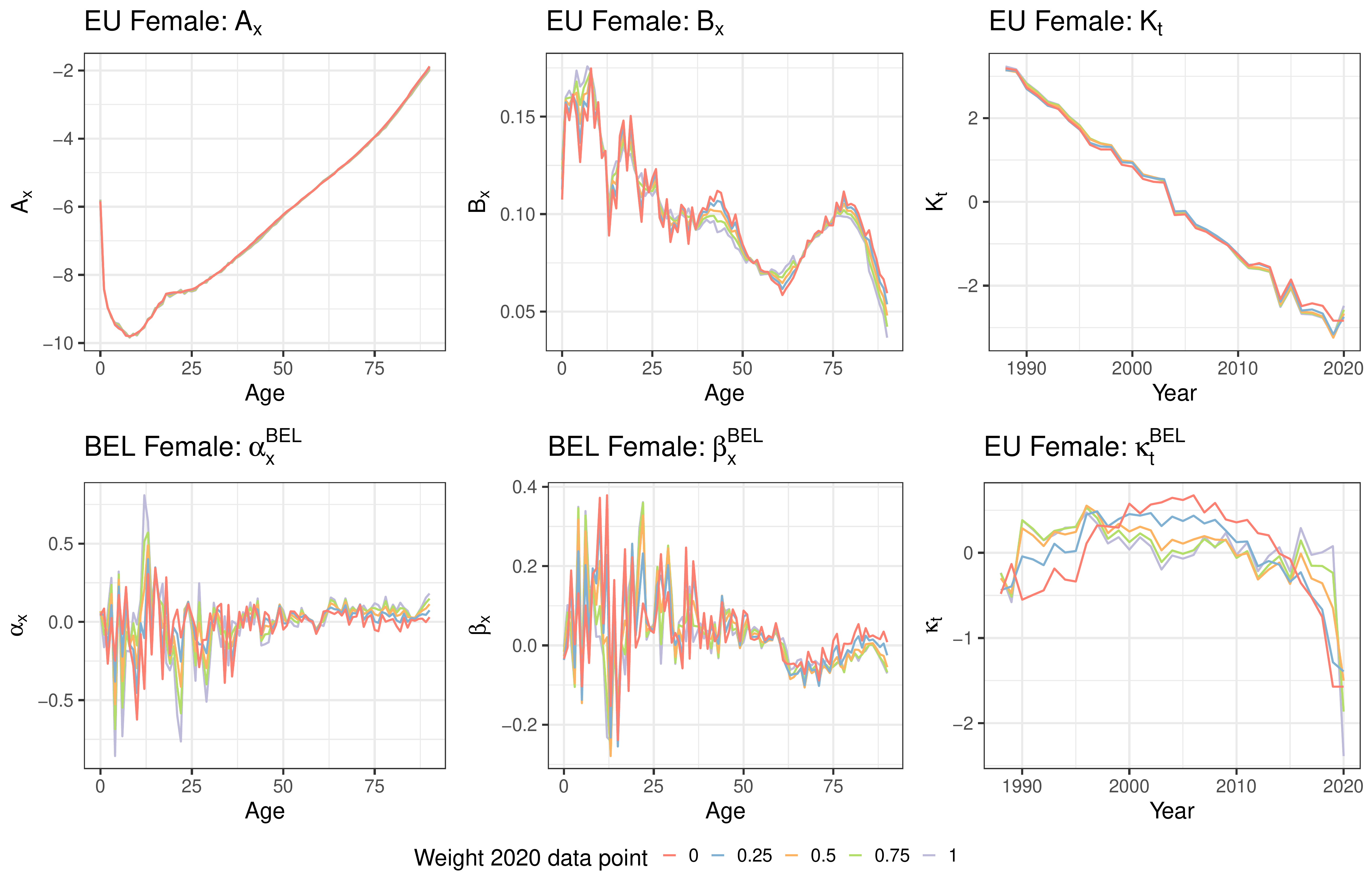}
\caption{The European and Belgian parameters $\hat{A}_x$, $\hat{B}_x$, $\hat{\alpha}_x$, $\hat{\beta}_x$, $\hat{K}_t$ and $\hat{\kappa}_t$ in the adjusted Lee \& Miller mortality model for males and females. The coloured lines correspond to the different weighting scenarios in the calibration set-up. The mortality model is calibrated on data from 1988 to 2020 and contains virtually created deaths and exposures for the years 2019-2020. \label{fig:eubelage_A2}}
\end{figure}

Figure~\ref{fig:eubelage_A2} displays the calibrated parameters for males and females in the adjusted Lee \& Miller mortality model. In contrast to Section~\ref{subsec:method1}, we now obtain different calibrated results for each weighting scenario. Note that the parameters $A_x$ and $\alpha_x$ are not calibrated in this approach, but they represent the fixed values, given in Equation~\eqref{eq:belmuxtMiller2}. In addition, the larger the weight $\alpha_{2020}$, the more pronounced the upward jump in the calibrated common period effects $\hat{K}_t$ and the larger the downward jump in the Belgium-specific period effects $\hat{\kappa}_t$ in the year 2020.

\begin{sidewaysfigure}
\begin{subfigure}{0.48\hsize}\centering
    \includegraphics[width=\hsize, height = 0.5\vsize]{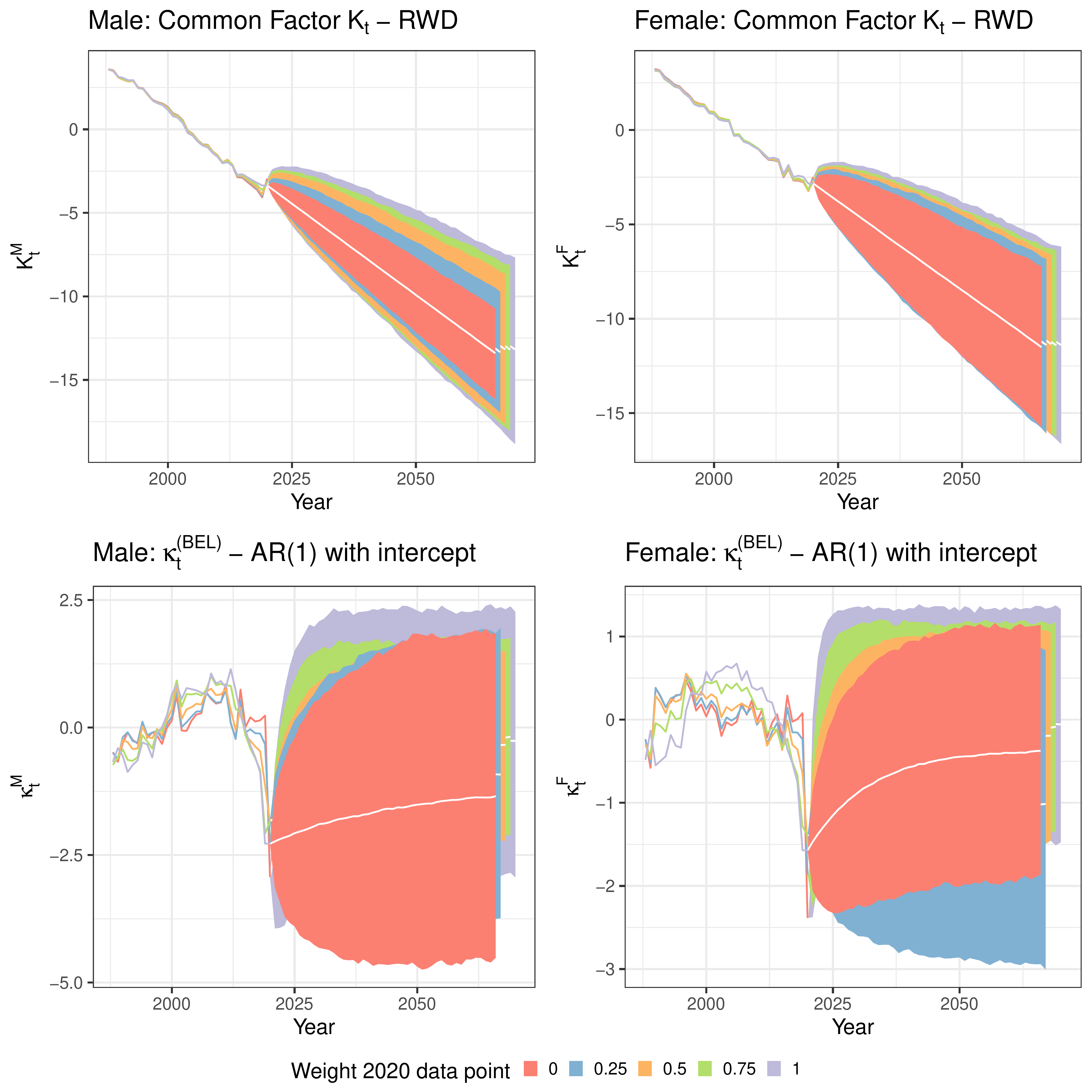}
\caption{Projection of the calibrated time dependent parameters in the adjusted Lee \& Miller model: $\hat{K}_t$ (top) and $\hat{\kappa}_t$ (bottom). Male (left) and female (right) data, calibration period $1988$-$2020$ and projection period $2021$-$2070$, method of Section~\ref{subsec:method2}. We show the $0.5\%$, median (white lines) and $99.5\%$ quantile, based on $10\ 000$ simulations across each weighting scenario. The coloured, solid lines visualize the calibrated period effects for each weighting scenario: $\hat{K}_t$ and $\hat{\kappa}_t$.{\color{white} white space white space white space white space white space white space white.} \label{fig:euKtmf_A2}}
\end{subfigure}%
\hfill 
\begin{subfigure}{0.48\hsize}\centering
    \includegraphics[width=\hsize, height = 0.5\vsize]{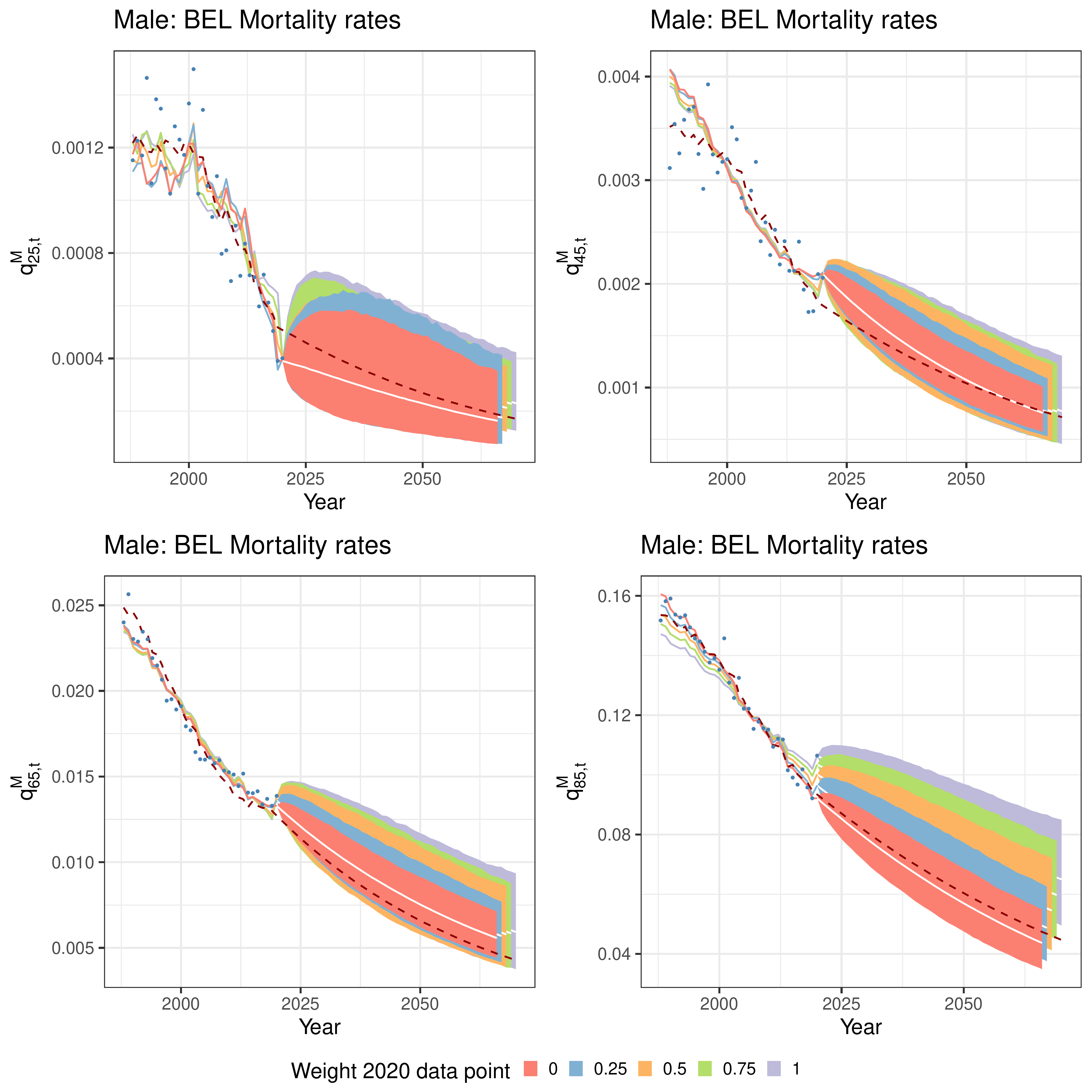}
\caption{Estimated and projected Belgian mortality rates $\hat{q}_{x,t}$. Male data, ages 25,45 (top) and 65, 85 (bottom), calibration period 1988-2020, projection period 2021-2070, method of Section~\ref{subsec:method2}. We show the $0.5\%$, median (white lines) and $99.5\%$ quantile (fan charts), based on $10\ 000$ simulations across each weighting scenario. The blue dots and the coloured, solid lines represent the observed and fitted mortality rates respectively. The dark red, dashed line shows the calibrated mortality rates and the median quantile of simulations in the original IA$\mid$BE 2020 model. \label{fig:qxtbelM25456585_A2}}
\end{subfigure}
\vspace{0.1cm}
\caption{Estimated and projected time dependent parameters (left) and Belgian mortality rates (right).}
\end{sidewaysfigure}

We use the same time dynamics and follow the same projection and simulation strategy as outlined in Sections~\ref{subsec:timedyna} and \ref{subsec:futurepaths}. We do not include weights in the time series likelihood (see Equation~\eqref{eq:4variatellweights}) to estimate the time series parameters. Figure~\ref{fig:euKtmf_A2} shows the calibrated and simulated period effects in the adjusted Lee \& Miller mortality model. Table~\ref{tab:modspec_A2} lists the estimated time series parameters.

\begin{table}[ht]
\centering
\begin{tabular}{lccccccc}
\toprule   
Weight 2020 &  $\theta^M$ & $\theta^F$ & $c^M$ & $c^F$ & $\phi^M$ & $\phi^F$  \\
\midrule
0    & -0.2179 & -0.1883 & -0.0472 & -0.0300 & 0.9605 & 0.9168 \\ 
0.25 & -0.2139 & -0.1839 & -0.0319 & -0.0284 & 0.9578 & 0.9680 \\
0.50 & -0.2096 & -0.1816 & -0.0333 & -0.0303 & 0.8990 & 0.8445 \\
0.75 & -0.2068 & -0.1800 & -0.0408 & -0.0318 & 0.7943 & 0.6746 \\
1    & -0.2048 & -0.1785 & -0.0562 & -0.0285 & 0.7786 & 0.4808 \\
\bottomrule
\end{tabular}
\caption{Time series parameter estimates, male and female data, ages $0$-$90$, years $1988$-$2020$, Method~\ref{subsec:method2}.\label{tab:modspec_A2}}
\end{table}

Figure~\ref{fig:qxtbelM25456585_A2} displays the calibrated and projected mortality rates for a 25, 45, 65 and 85 year old Belgian male. A lower weight $\alpha_{2020}$ implies that the fitted mortality rates in 2020 are closer to the observed mortality rates in 2019 than those of 2020. This in turn results in an overall, better in-sample fit of the mortality rates at old ages, see e.g.~the bottom right panel in Figure~\ref{fig:qxtbelM25456585_A2}. Moreover, lowering the weight $\alpha_{2020}$ leads to results closer to the original IA$\mid$BE 2020 model (dark red, dashed line) on average. Another noteworthy fact is the increase in the projected mortality rates for a 25 year old male (top, left panel) right after the pandemic year 2020 in the case of a larger weight $\alpha_{2020}$. This increase is partly due the fast mean reversion of the time series (see Figure~\ref{fig:euKtmf_A2}) in combination with larger calibrated $\hat{B}_x$ values at the younger ages (see Figure~\ref{fig:eubelage_A2} for the graphs of $\hat{B}_x$). Therefore, the term $\beta_x \cdot (\kappa_t - \kappa_{2020})$ can become larger than the term $B_x\cdot(K_t - K_{2020})$ in Equation~\eqref{eq:belmuxtMiller2} at early years in the projection period.

\begin{figure}[h!]
\centering
\includegraphics[width = 0.8\textwidth]{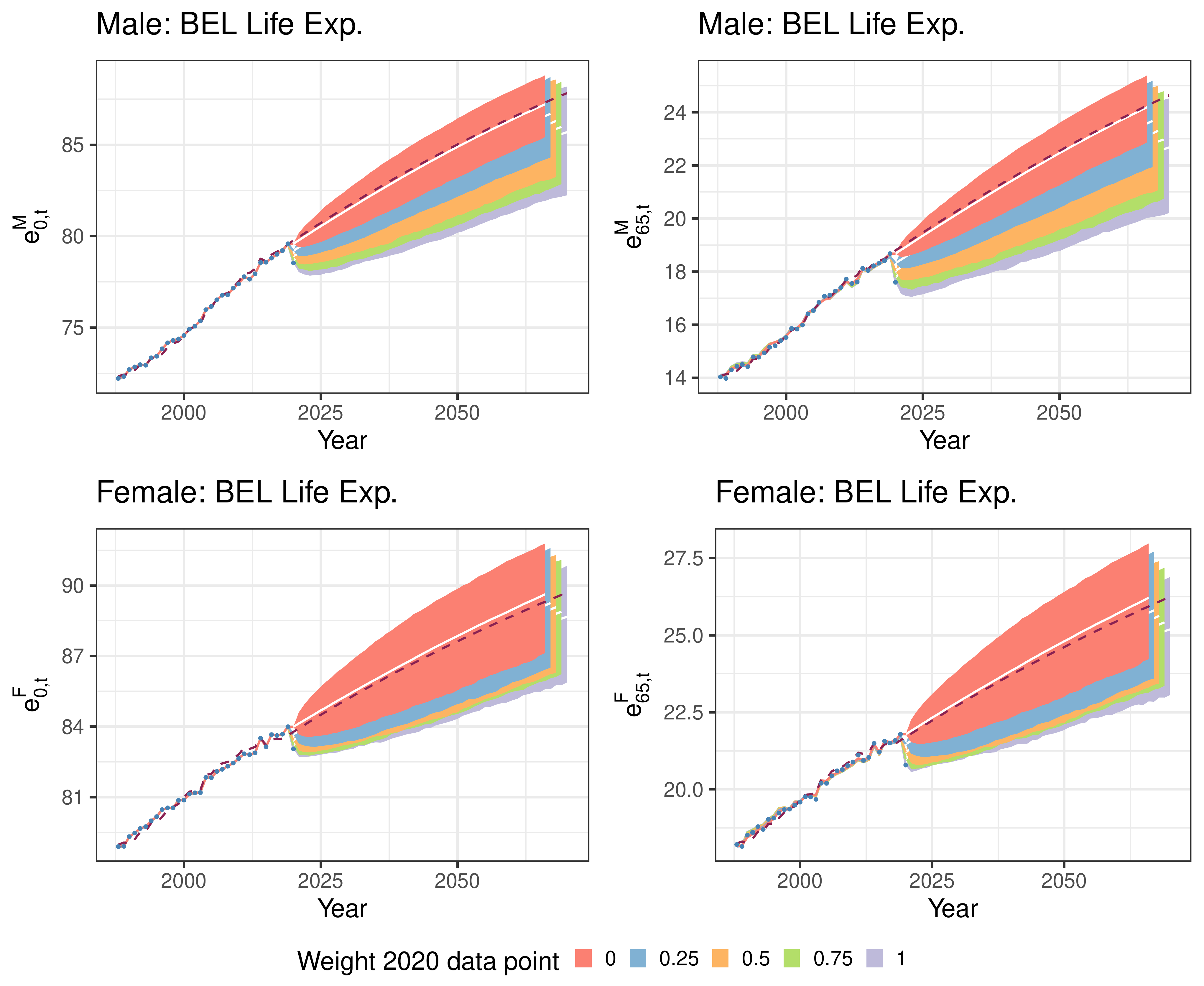}
\caption{Estimated and projected period life expectancies $\hat{e}_{x,t}$ for Belgium. Male (top) and female (bottom) data, ages 0 (left) and 65 (right), calibration period 1988-2020, projection period 2021-2070. Calibration and projection is based on the method in Section~\ref{subsec:method1}. We show the $0.5\%$ quantile, median (white lines) and $99.5\%$ quantile (fan charts), based on $10\ 000$ simulations, across each weighting scenario. The dark red, dashed line shows the $50\%$ quantile originating from the original IA$\mid$BE 2020 model. The blue dots represent the observed period life expectancies.\label{fig:extbelMF065_A2}}
\end{figure}

Figure~\ref{fig:extbelMF065_A2} shows the estimated and projected period life expectancies for a 0 and 65 year old Belgian male (top) and female (bottom). Assigning the value zero to the weight $\alpha_{2020}$ leads to comparable results with the original IA$\mid$BE 2020 model (dark red, dashed line). This is in line with our expectations since we actually omit the pandemic data point in the calibration step when $\alpha_{2020} = 0$. Table~\ref{tab:A2CLE065} depicts the cohort life expectancy in 2020 for a 0 and 65 year old male and female.

\begin{table}[ht]
\centering
\adjustbox{max width=\textwidth}{%
\centering
\begin{tabular}{@{\extracolsep{4pt}}lccccc}
\toprule   
\multicolumn{2}{l}{\multirow{2}{4cm}{\textbf{Cohort life expectancy in 2020}}}  & \multicolumn{2}{c}{Males}  & \multicolumn{2}{c}{Females}\\
 \cmidrule{3-4} 
 \cmidrule{5-6} 
 & & 0 & 65 & 0 &65 \\
\midrule
\multirow{2}{*}{\shortstack[l]{Recalibration \\ \scriptsize{2020 weight = 0}}} 	& Best. Est.    & 89.91 & 20.30 & 92.11 & 23.32 \\
 &$[q_{0.5};q_{50};q_{99.5}]$ & $[88.27;89.91;91.37]$ & $[19.57;20.30;21.04]$ & $[89.53;92.11;94.15]$ & $[22.16;23.31;24.42]$ \\ \\
 \multirow{2}{*}{\shortstack[l]{Recalibration \\ \scriptsize{2020 weight = 0.25}}} 	& Best. Est.   & 89.01 & 19.83 & 91.43 & 22.94 \\
 &$[q_{0.5};q_{50};q_{99.5}]$ & $[86.81;88.98;90.74]$ & $[18.95;19.82;20.68]$ & $[88.66;91.41;93.62]$ & $[21.70;22.94;24.16]$\\ \\
\multirow{2}{*}{\shortstack[l]{Recalibration \\ \scriptsize{2020 weight = 0.50}}} 	& Best. Est.   & 88.22 & 19.5 & 90.94 & 22.79 \\
 &$[q_{0.5};q_{50};q_{99.5}]$ & $[85.52;88.22;90.20]$ & $[18.38;19.49;20.52]$ & $[88.22;90.94;92.96]$ & $[21.51;22.78;23.97]$\\ \\
\multirow{2}{*}{\shortstack[l]{Recalibration \\ \scriptsize{2020 weight = 0.75}}} 	& Best. Est.   & 87.61 & 19.31 & 90.49 & 22.66 \\
 &$[q_{0.5};q_{50};q_{99.5}]$ & $[84.77;87.59;89.73]$ & $[18.09;19.30;20.46]$ & $[87.97;90.49;92.50]$ & $[21.39;22.66;23.79]$\\ \\
\multirow{2}{*}{\shortstack[l]{Recalibration \\ \scriptsize{2020 weight = 1}}} 	    & Best. Est.   & 87.07 & 19.10 & 90.06 & 22.52 \\
 &$[q_{0.5};q_{50};q_{99.5}]$ & $[83.98;87.04;89.22]$ & $[17.85;19.08;20.27]$ & $[87.47;90.03;91.94]$ & $[21.34;22.50;23.6]$\\ \\
\multirow{2}{*}{IA$\mid$BE 2020} & Best. Est.   & 89.91 & 20.38 & 91.54 & 23.14\\
 &$[q_{0.5};q_{50};q_{99.5}]$ & $[88.11;89.89;91.46]$ & $[19.57;20.37;21.17]$ & $[89.46;91.53;93.25]$ & $[22.15;23.14;24.07]$\\ \\
\bottomrule
\end{tabular}}
\caption{The cohort life expectancy for a 0 and 65 year old in 2020. The best estimate and the $0.5\%$ quantile, median and $99.5\%$ quantile obtained from 10\ 000 simulations are shown, for males and females.\label{tab:A2CLE065}}
\end{table} 

\section{Conclusion and outlook} \label{sec:outlook}
This paper examines different methods to deal with a pandemic data point in the calibration and projection set-up of a stochastic multi-population mortality projection model, in casu the Li \& Lee model. When this data point corresponds to the last observed year in the calibration period, it severely affects the drift parameter estimation in the random walk with drift process to model the common European period effect for males and females. To mitigate this impact we propose to make changes in either the projection strategy (Section~\ref{subsec:method1}) or in the model specifications (Section~\ref{subsec:method2}) itself. We do this by restricting the impact of the pandemic year through the inclusion of weights in the calibration or projection step.

There are still many uncertainties about the future evolution of COVID-19. COVID-19 may have impacted mortality rates during just one year, or future years may be affected as well. Future relevant work may focus on modifying the Li \& Lee model so that it can automatically absorb pandemic shocks rather than assigning a subjective weight to this pandemic data point. \citet{schnurch2021impact} mention some first ideas to handle extreme mortality events including outlier analysis, regime switching models, the use of techniques from extreme value theory and the use of jump processes in the time series model for the period effect.

In addition, several assumptions underneath the stochastic multi-population mortality projection model of type Li \& Lee may require further investigation. Future work may put focus on including a cohort effect in the model. This can be useful when COVID-19 has a long-lasting effect on the health of people who have been severely affected by COVID-19, e.g.~hospitalized persons. A second research topic is to select the weight $\alpha_{2020}$, assigned to the observed death rates in 2020 in Section~\ref{subsec:method2}, in a data-driven way. One idea is to include $\alpha_{2020}$ as a parameter that can be optimally chosen in the calibration set-up. Finally, future research may focus on the performance of Kannisto's method to extrapolate the mortality rates above the age of 90 in the presence of a pandemic shock. 

\section*{Acknowledgements}
Katrien Antonio acknowledges the support of the research chair DIALog, sponsored by CNP Assurances.

\appendix

\bibliographystyle{plainnat}
\bibliography{Ref_Covid}

\begin{thebibliography}{19}
\providecommand{\natexlab}[1]{#1}
\providecommand{\url}[1]{\texttt{#1}}
\expandafter\ifx\csname urlstyle\endcsname\relax
  \providecommand{\doi}[1]{doi: #1}\else
  \providecommand{\doi}{doi: \begingroup \urlstyle{rm}\Url}\fi

\bibitem[Antonio et~al.(2017)Antonio, Devriendt, de~Boer, de~Vries,
  De~Waegenaere, Kan, Kromme, Ouburg, Schulteis, Slagter,
  et~al.]{antonio2017producing}
Katrien Antonio, Sander Devriendt, Wouter de~Boer, Robert de~Vries, Anja
  De~Waegenaere, Hok-Kwan Kan, Egbert Kromme, Wilbert Ouburg, Tim Schulteis,
  Erica Slagter, et~al.
\newblock Producing the {D}utch and {B}elgian mortality projections: a
  stochastic multi-population standard.
\newblock \emph{European actuarial journal}, 7\penalty0 (2):\penalty0 297--336,
  2017.
\newblock \doi{10.1007/s13385-017-0159-x}.

\bibitem[Antonio et~al.(2020)Antonio, Devriendt, and Robben]{IABE2020}
Katrien Antonio, Sander Devriendt, and Jens Robben.
\newblock The {IA$\mid$BE} 2020 mortality projection for the {B}elgian
  population, 2020.

\bibitem[B{\"o}rger et~al.(2014)B{\"o}rger, Fleischer, and
  Kuksin]{borger2014modeling}
Matthias B{\"o}rger, Daniel Fleischer, and Nikita Kuksin.
\newblock Modeling the mortality trend under modern solvency regimes.
\newblock \emph{ASTIN Bulletin: The Journal of the IAA}, 44\penalty0
  (1):\penalty0 1--38, 2014.
\newblock \doi{10.1017/asb.2013.24}.

\bibitem[Brouhns et~al.(2002)Brouhns, Denuit, and Vermunt]{BrouhnsDenuit}
Natacha Brouhns, Michel Denuit, and Jeroen~K Vermunt.
\newblock A {P}oisson log-bilinear regression approach to the construction of
  projected lifetables.
\newblock \emph{Insurance: Mathematics and economics}, 31\penalty0
  (3):\penalty0 373--393, 2002.
\newblock \doi{10.1016/S0167-6687(02)00185-3}.

\bibitem[Cairns et~al.(2009)Cairns, Blake, Dowd, Coughlan, Epstein, Ong, and
  Balevich]{CairnsNAAJ}
A.J.G. Cairns, D.~Blake, K.~Dowd, G.D. Coughlan, D.~Epstein, A.~Ong, and
  I.~Balevich.
\newblock A quantitative comparison of stochastic mortality models using data
  from {E}ngland and {W}ales and the {U}nited {S}tates.
\newblock \emph{North American Actuarial Journal}, 13\penalty0 (1):\penalty0
  1--35, 2009.
\newblock \doi{10.1080/10920277.2009.10597538}.

\bibitem[Genootschap(2018)]{KAG2018}
Koninklijk~Actuarieel Genootschap.
\newblock Prognosetafel {AG}2018.
\newblock \url{https://www.ag-ai.nl}, 2018.

\bibitem[Genootschap(2020)]{KAG2020}
Koninklijk~Actuarieel Genootschap.
\newblock Prognosetafel {AG}2020.
\newblock \url{https://www.ag-ai.nl}, 2020.

\bibitem[Haberman and Renshaw(2011)]{HabRenshIME2011}
S.~Haberman and A.E. Renshaw.
\newblock A comparative study of parametric mortality models.
\newblock \emph{Insurance: Mathematics and Economics}, 48\penalty0
  (1):\penalty0 35--55, 2011.
\newblock \doi{10.1016/j.insmatheco.2010.09.003}.

\bibitem[Haberman et~al.(2014)Haberman, Kaishev, Millossovich, Villegas,
  Baxter, Gaches, Gunnlaugsson, and Sison]{haberman2014longevity}
Steven Haberman, Vladimir Kaishev, Pietro Millossovich, Andr{\'e}s Villegas,
  Steven Baxter, Andrew Gaches, Sveinn Gunnlaugsson, and Mario Sison.
\newblock Longevity basis risk: A methodology for assessing basis risk.
\newblock Technical report, Institute and Faculty of Actuaries, 2014.

\bibitem[Kannisto(1994)]{Kannisto}
V{\"a}in{\"o} Kannisto.
\newblock \emph{Development of oldest-old mortality, 1950-1990: {E}vidence from
  28 developed countries}.
\newblock Odense University Press, 1994.

\bibitem[Lee and Miller(2001)]{leemiller}
Ronald Lee and Timothy Miller.
\newblock Evaluating the performance of the {L}ee-{C}arter method for
  forecasting mortality.
\newblock \emph{Demography}, 38\penalty0 (4):\penalty0 537--549, 2001.
\newblock \doi{10.1353/dem.2001.0036}.

\bibitem[Lee and Carter(1992)]{LeeCarter}
Ronald~D. Lee and Lawrence~R. Carter.
\newblock Modeling and forecasting {U.S.} mortality.
\newblock \emph{Journal of the American Statistical Association}, 87\penalty0
  (419):\penalty0 659--671, 1992.
\newblock \doi{10.2307/2290201}.

\bibitem[Li(2013)]{Li2013}
Jackie Li.
\newblock A {P}oisson common factor model for projecting mortality and life
  expectancy jointly for females and males.
\newblock \emph{Population Studies: A Journal of Demography}, 67\penalty0
  (1):\penalty0 111--126, 2013.
\newblock \doi{10.1080/00324728.2012.689316}.
\newblock PMID: 22788919.

\bibitem[Li and Lee(2005)]{LiandLee}
N.~Li and R.~Lee.
\newblock Coherent mortality forecasts for a group of populations: an extension
  of the {L}ee--{C}arter method.
\newblock \emph{Demography}, 42\penalty0 (3):\penalty0 575--594, 2005.
\newblock \doi{10.1353/dem.2005.0021}.

\bibitem[Pitacco et~al.(2009)Pitacco, Denuit, Haberman, and Olivieri]{pitacco}
E.~Pitacco, M.~Denuit, S.~Haberman, and A.~Olivieri.
\newblock \emph{Modeling Longevity Dynamics for Pensions and Annuity Business}.
\newblock Oxford University Press, London, 2009.

\bibitem[Rizzi et~al.(2015)Rizzi, Gampe, and Eilers]{rizzi2015efficient}
Silvia Rizzi, Jutta Gampe, and Paul~HC Eilers.
\newblock Efficient estimation of smooth distributions from coarsely grouped
  data.
\newblock \emph{American Journal of Epidemiology}, 182\penalty0 (2):\penalty0
  138--147, 2015.
\newblock \doi{10.1093/aje/kwv020}.

\bibitem[Schn{\"u}rch et~al.(2021)Schn{\"u}rch, Kleinow, Korn, and
  Wagner]{schnurch2021impact}
Simon Schn{\"u}rch, Torsten Kleinow, Ralf Korn, and Andreas Wagner.
\newblock The impact of mortality shocks on modeling and insurance valuation as
  exemplified by {COVID}-19 ({S}eptember 29, 2021).
\newblock Available at SSRN: \url{https://dx.doi.org/10.2139/ssrn.3835907},
  2021.

\bibitem[Van~Berkum et~al.(2016)Van~Berkum, Antonio, and
  Vellekoop]{VanBerkum2014}
Frank Van~Berkum, Katrien Antonio, and Michel Vellekoop.
\newblock The impact of multiple structural changes on mortality predictions.
\newblock \emph{Scandinavian Actuarial Journal}, 2016\penalty0 (7):\penalty0
  581--603, 2016.
\newblock \doi{10.1080/03461238.2014.987807}.

\bibitem[van Delft and Huijzer(2020)]{milliman}
Lotte van Delft and Sarah Huijzer.
\newblock Impact of {COVID}-19 on {D}utch mortality tables.
\newblock
  \url{https://be.milliman.com/-/media/milliman/pdfs/articles/impact-of-covid-19-on-dutch-mortality-tables.ashx},
  2020.

\end{thebibliography}
\appendix
\section{Data sources}\label{sec:overview}
\begin{table}[ht]
\centering
\begin{tabular}{@{\extracolsep{4pt}}lcccccccc}
\toprule   
{}  & \multicolumn{4}{c}{\textbf{Exposures}}  & \multicolumn{4}{c}{\textbf{Deaths}}\\
 \cmidrule{2-5} 
 \cmidrule{6-9} 
Country & 2017 & 2018 & 2019 & 2020 & 2017 & 2018 & 2019 & 2020 \\ 
\midrule
AUS  & HMD & HMD & HMD & STMF & HMD & HMD & HMD & EURO.W \\ 
BEL  & HMD & HMD  & EURO & STMF & HMD & HMD  & EURO   & STATBEL \\ 
DNK  & HMD & HMD  & HMD  & HMD & HMD & HMD  & HMD    & HMD \\ 
FIN  & HMD & HMD  & HMD  & STMF & HMD & HMD  & HMD    & EURO.W \\
FRA  & HMD & HMD  & EURO & STMF & HMD & HMD  & EURO   & STMF \\ 
GER  & HMD & EURO & EURO & STMF & HMD & EURO & EURO   & STMF \\ 
ICE  & HMD & HMD  & EURO & STMF & HMD & HMD  & EURO & EURO.W \\ 
LUX  & HMD & HMD  & HMD  & STMF & HMD & HMD  & HMD    & EURO.W \\ 
NED  & HMD & HMD  & HMD & STMF & HMD & HMD  & HMD & EURO.W \\ 
NOR  & HMD & HMD  & EURO & STMF & HMD & HMD  & EURO & EURO.W \\ 
SWE  & HMD & HMD  & HMD  & STMF & HMD & HMD  & HMD    & EURO.W \\ 
SWI  & HMD & HMD  & EURO & STMF & HMD & HMD  & EURO & EURO.W \\ 
UNK  & HMD & HMD  & STMF & STMF & HMD & HMD  & STMF   & STMF \\ 
\bottomrule
\end{tabular}
\caption{Overview of the data sources used for each country in the stochastic multi-population mortality projection model of type Li \& Lee. The data sources `HMD', `EURO' and `STATBEL' refer to the Human Mortality Database, Eurostat and the Belgian statistical institute Statbel respectively. They provide mortality data on an annual basis and at the level of individual ages. `HMD' is our primary data source. We use the other two data sources to supplement the annual deaths and exposures at individual age level for the more recent years 2018-2020 where possible. Because these data sources are subject to a significant reporting delay of sometimes several years (e.g.~for the United Kingdom), we consult the Short-Term Mortality Fluctuations (STMF) Data series and the weekly death statistics available at Eurostat (EURO.W) to supplement our dataset until the year 2020. The latter two data sources provide weekly mortality statistics registered in age buckets. We convert these to annual mortality statistics at the individual age level using the protocol in Section~\ref{sec:virtdata1920}.\label{tab:overview} } 
\end{table}

\section{Constructing virtual exposure points} \label{sec:create.exp}
We create virtual annual exposures $E_{x,t}$ for individual ages 0-90, years 2019-2020 and for each country that is included in the calibration of the common, multi-population trend in the Li \& Lee mortality projection model (Section~\ref{subsec:mortprojbe}).\footnote{For all countries except the United Kingdom, we do have the annual exposures $E_{x,t}$ at individual age level in 2019 from either HMD or Eurostat (see Table~\ref{tab:overview}). For Denmark, we already have the annual exposures $E_{x,t}$ at individual age level in the year 2020 available from the HMD.} We explain our strategy to create virtual exposures for Belgium in the year 2020, but we follow a similar approach for any other country that is part of this common mortality trend. 

\begin{figure}[h!]
\centering
\includegraphics[width = 0.8\textwidth]{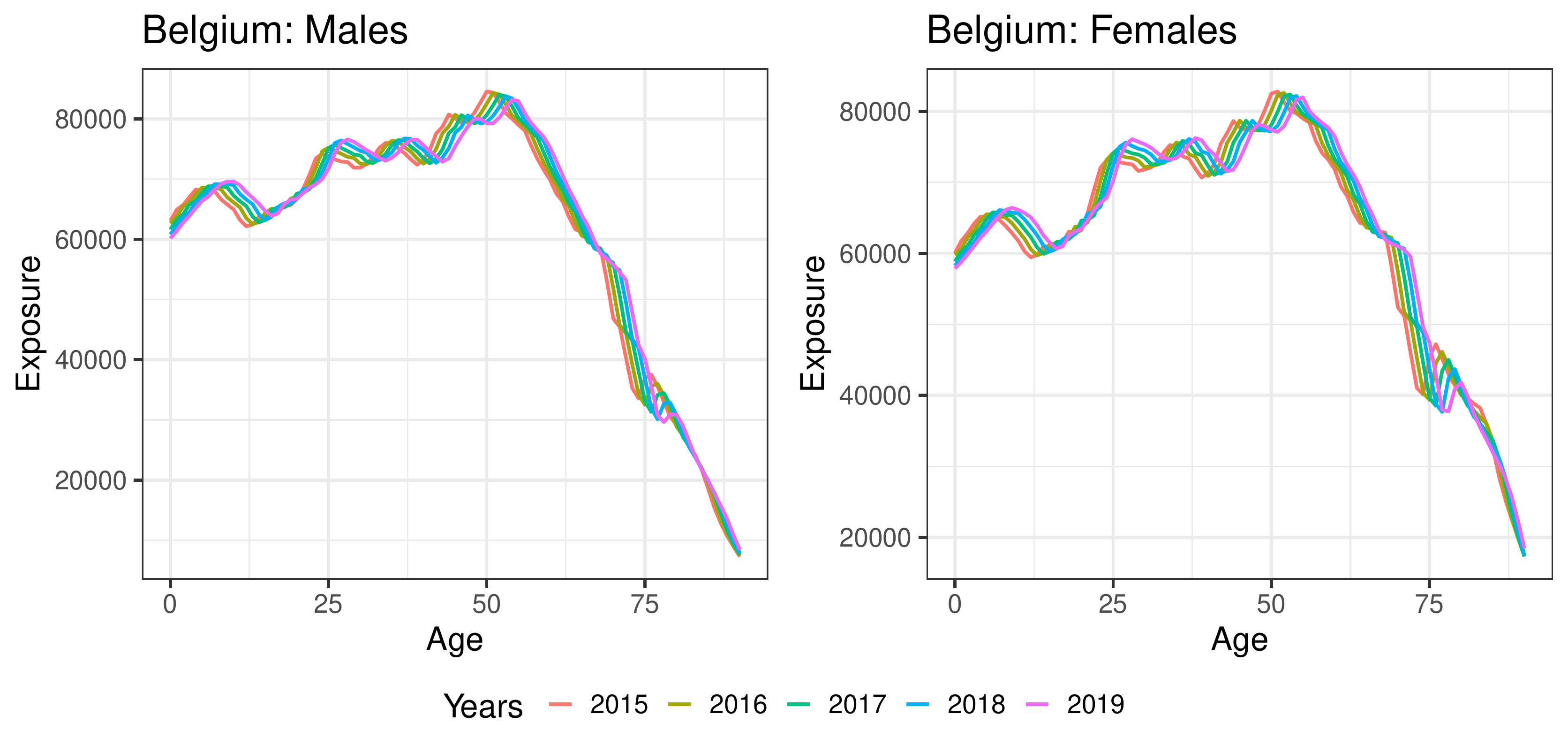}
\caption{The exposures $E_{x,t}$ of Belgium for ages 0-90 and years 2015-2019. Data from HMD until the year 2018 and from Eurostat for the year 2019.\label{fig:Extbel1519}}
\end{figure}

Figure~\ref{fig:Extbel1519} shows the observed exposures in Belgium as a function of age over the years 2015-2019. These exposures are retrieved from HMD and Eurostat (see Table~\ref{tab:overview}). The exposure function has a similar pattern shifted to the right with one age in each subsequent year $t$. This is in line with our intuition since people aged $x$ in year $t$ become part of the group at risk aged $x+1$ in year $t+1$, in case of survival.

\begin{figure}[h!]
\centering
\includegraphics[width = 0.8\textwidth]{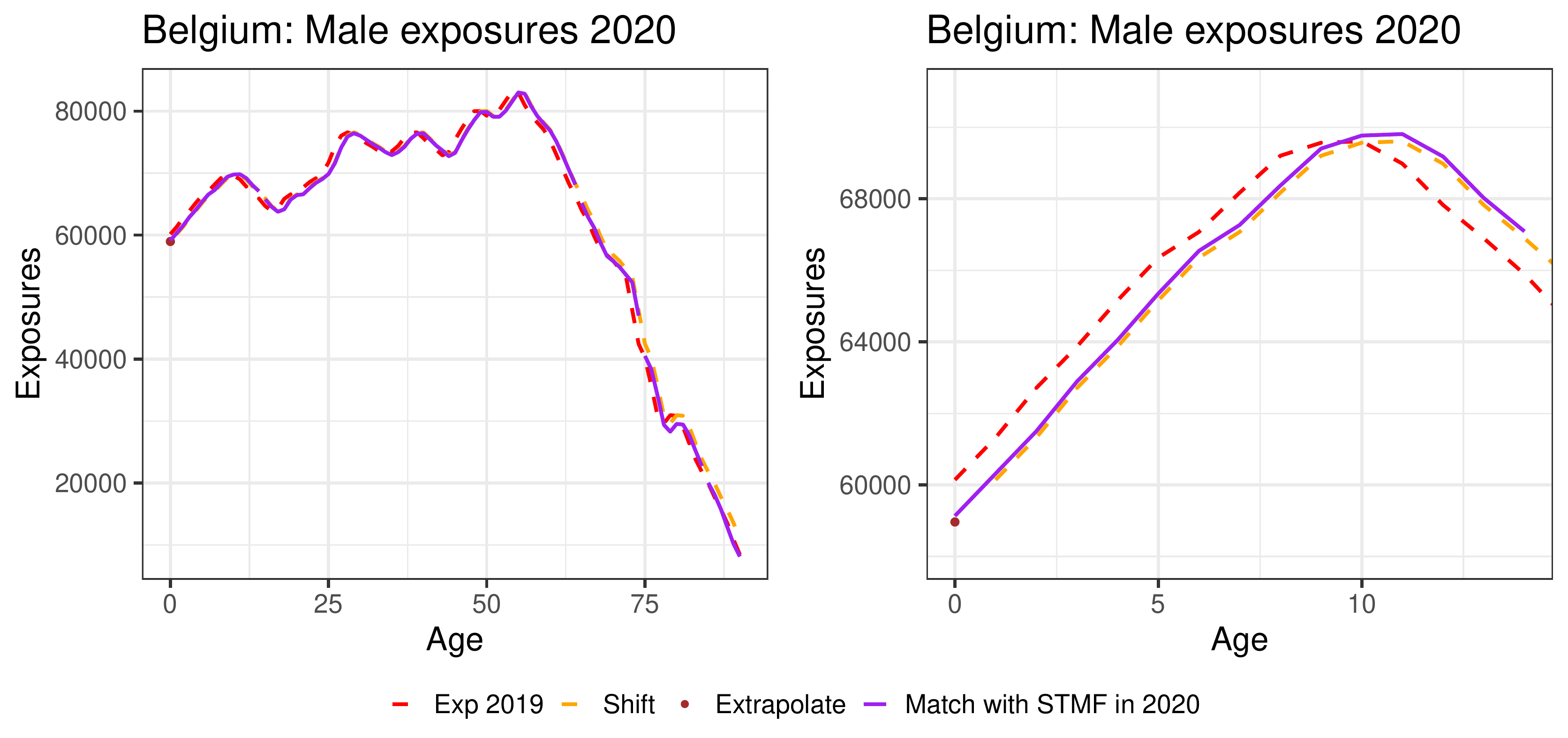}
\caption{The virtual exposure $E_{x,t}$ for Belgium in the year $2020$, for males. At the right, we show a snapshot for the age range 0-14. \label{fig:Extbel2020}}
\end{figure}

We use the above reasoning to create virtual annual exposures for $2020$. Figure~\ref{fig:Extbel2020} graphically explains the strategy for Belgian males. First, we start from the observed annual exposures $E_{x,t}$ in the most recent available year (in casu 2019 for Belgium, using Eurostat) and visualize these as a function of age (red, dashed line). Second, we shift this 2019 exposure curve one age to the right, resulting in the orange, dashed line. This newly created curve is undefined at the age of zero because of the shift to the right. Therefore, in a third step, we linearly extrapolate the orange, dashed line to zero.\footnote{We use the exposure points $(1,E_{1,t})$ and $(2,E_{2,t})$ to linearly extrapolate to age $0$: $$E_{0,t} = E_{1,t} + \dfrac{E_{2,t} - E_{1,t}}{2 - 1}\cdot(0-1).$$} This choice is justified by the linear pattern of the exposure function at young ages. We obtain the brown exposure point in Figure~\ref{fig:Extbel2020}. We denote the orange, dashed line extended with the brown point at age $0$ as $\hat{E}_{x,t}^s$. In a last step, we match the new exposure function of 2020, i.e.~$\hat{E}_{x,2020}^s$, with the exposures collected in age buckets from the STMF data series, as shown in Table~\ref{tab:expbucketbel}. Hereto, we consider an age bucket $[x_i,x_j]$ and define the virtual annual exposures $E_{x,t}$ as:
\begin{equation}\label{eq:exptransition}
\begin{aligned}
E_{x,t} = \hat{E}_{x,t}^s \cdot b_{i,j}, \hspace{0.5cm} \text{where} \: \: \:
b_{i,j} = \dfrac{E_{[x_i,x_j],t}}{\displaystyle \sum_{a = x_i}^{x_j} \hat{E}_{a,t}^s},
\end{aligned}
\end{equation}
for $t = 2020$ (for example) and $x \in [x_i,x_j]$. Intuitively, we vertically scale a section of the orange dashed line, corresponding to a certain age bucket, such that the summed exposure within this age bucket corresponds to the total exposure in the same age bucket of Table~\ref{tab:expbucketbel}. The right panel of Figure~\ref{fig:Extbel2020} shows this strategy for the age bucket $[0,14]$, where the purple line shows the final virtual exposures $E_{x,t}$ at individual ages for Belgium in the year $2020$. 
\begin{table}[h!]
\centering
\begin{tabular}{@{\extracolsep{4pt}}lcc}
\toprule   
Age bucket  & \multicolumn{1}{c}{Male Exp.}  & \multicolumn{1}{c}{Female Exp.}\\
\midrule 
$[0,14]$ & 988\ 713.02 & 944\ 379.40 \\ 
$[15,64]$ & 3\ 699\ 434.72 & 3\ 638\ 808.41 \\ 
$[65,74]$ & 568\ 101.96 & 618\ 244.99 \\ 
$[75,84]$ & 305\ 175.72 & 399\ 015.96 \\ 
$85+$ & 112\ 577.56 & 223\ 565.55 \\ 
   \hline
\end{tabular}
\caption{The male and female Belgian exposures in 2020 in age buckets, obtained from the STMF data series.\label{tab:expbucketbel}}
\end{table}

We apply a slightly different strategy for the exposure $E_{85+, 2020}$ reported for ages in the open age bucket $85+$ on the STMF data series. The underlying idea is that we want to distribute the extra exposure $E_{85+, 2020}$ (from the STMF data series) minus $E_{85+, 2019}$ (from HMD) evenly across the ages $85+$, i.e.~ages $85,\ldots, 110$ (the assumed maximum age). Hereto, we calculate the shift $c_{85+}$, as follows
\begin{equation*}
\begin{aligned}
c_{85+} = \dfrac{1}{110-85+1} \left(E_{85+,2020} - E_{85+, 2019}\right).
\end{aligned}
\end{equation*}
We then apply this shift $c_{85+}$ to go from $E_{x,2019}$ to $E_{x,2020}$: 
\begin{align*}
E_{x,2020} = E_{x,2019} + c_{85+},
\end{align*}
for ages $x \in \{85,86,\ldots, 90\}$.

We repeat this procedure for all 13 European countries. In case there is no exposure data available for 2019 on the HMD (see Table~\ref{tab:overview} in Appendix~\ref{sec:overview}), we start from the exposure curve for 2019 reported on the HMD or Eurostat and repeat the procedure two times to generate $E_{x,t}$ data points for the years 2019-2020.
 
\section{Constructing virtual death counts}\label{sec:create.deathcounts}
We construct virtual annual death counts $d_{x,t}$ at individual ages 0-90, years 2019-2020 and for each country that is included in the calibration of the common, multi-population trend in the Li \& Lee mortality projection model (Section~\ref{subsec:mortprojbe}).\footnote{For all countries except the United Kingdom, we do have the annual death counts $d_{x,t}$ at individual ages in 2019 from either HMD or Eurostat (see Table~\ref{tab:overview}). For Belgium and Denmark we even have the annual death counts $d_{x,t}$ at individual age level in 2020 available from HMD and Statbel respectively. For these two countries, there is no need to create virtual death counts.} In explaining our strategy, we focus on constructing virtual 2020 death counts at individual age level for the Netherlands, but a similar approach can be taken for any other country that is part of this common mortality trend. 

\begin{figure}[h!]
\centering
\includegraphics[width = 0.8\textwidth]{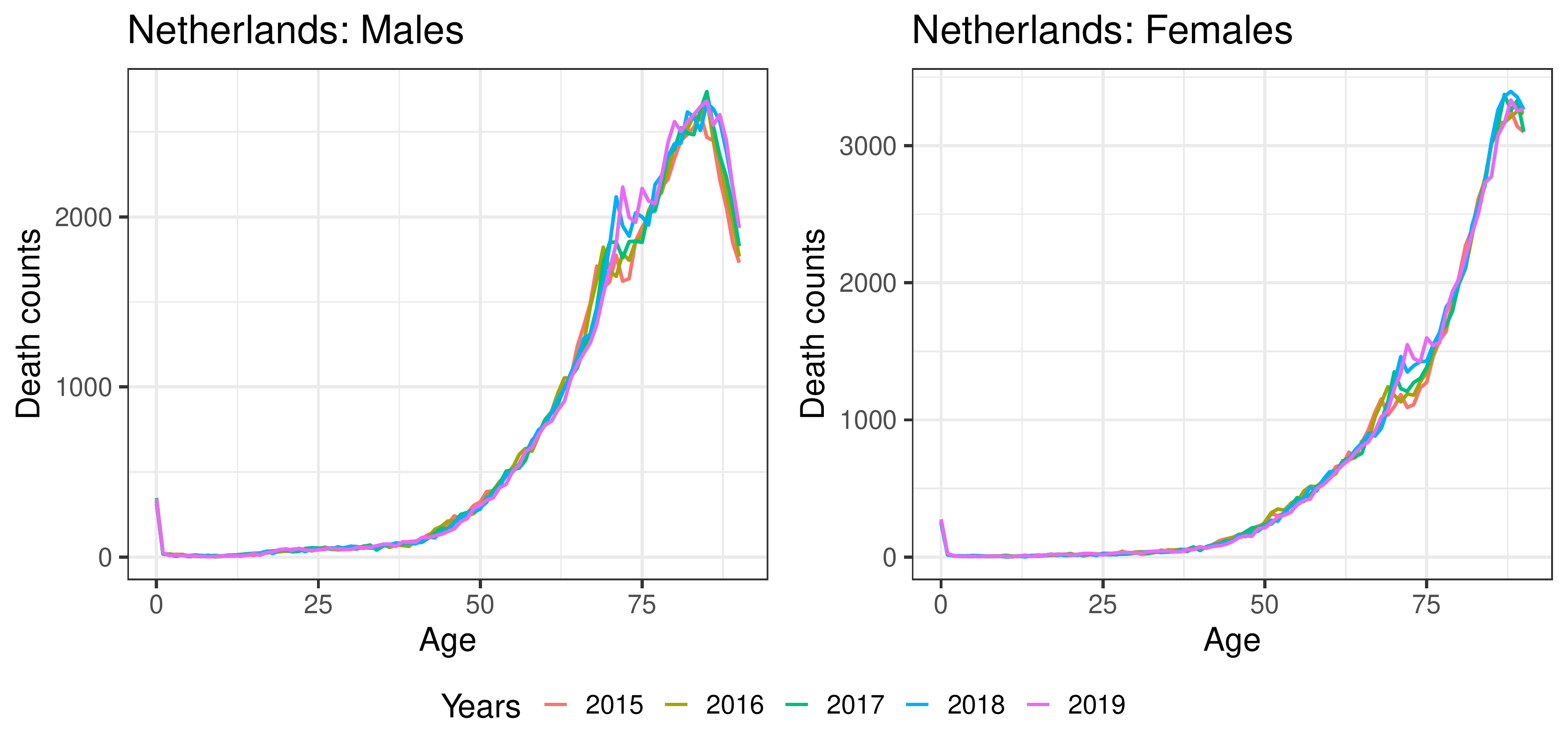}
\caption{The death counts $d_{x,t}$ of the Netherlands for ages 0-90 and years 2015-2019. Data from HMD.\label{fig:dxtnld1519}}
\end{figure}

Figure~\ref{fig:dxtnld1519} shows the observed annual deaths in the Netherlands across ages 0-90 and over the years 2015-2019 for males (left) and females (right). In line with our discussion about the pattern of the exposure curve in Figure~\ref{fig:Extbel1519}, we observe a time-effect in these death counts, e.g.~the bumps in the deaths pattern move to the right each consecutive year. We keep this in mind to construct virtual deaths for the year 2020.

We propose the following strategy. Using a Li \& Lee mortality model that is calibrated only on the observed annual deaths and exposures from HMD and/or Eurostat, we project the fitted force of mortality $\hat{\mu}_{x,t}^{\text{c}}$ over the next year(s) (see Section~\ref{subsec:futurepaths}). Similar to the discussion in \citet{IABE2020}, we choose the starting year of the calibration period in the Li \& Lee mortality model such that we retrieve stable AR(1) processes for both the male and female Dutch period effect. This motivates the use of starting year 1970 for the case of the Netherlands. Under the assumption of a piecewise constant force of mortality, the maximum likelihood estimate of the force of mortality $\mu_{x,t}^{\text{MLE,c}}$ then equals
\begin{align}\label{eq:mutod}
\hat{\mu}_{x,t}^{\text{MLE,c}} = m_{x,t}^{\text{c}} = \dfrac{d_{x,t}^{\text{c}}}{E_{x,t}^{\text{c}}},
\end{align}
with $c$ the country of interest, i.e.~the Netherlands in the example under consideration.

Note that we create virtual annual exposures $E_{x,t}$ for the Netherlands in the year 2020 according to the strategy explained in Appendix~\ref{sec:create.exp}. Using Equation~\eqref{eq:mutod}, we can then easily make the transition to virtual death counts $\hat{d}_{x,t}^{\text{c}}$. In a next step we match these expected deaths $\hat{d}_{x,t}^{\text{c}}$ with the information we retrieve from the weekly deaths data on Eurostat or the STMF data series. For the Netherlands, we work with the weekly death counts in age buckets of length 5 from Eurostat, see Table~\ref{tab:firsthalf2020}. 
\begin{table}[ht]
    \centering
    \begin{tabular}{lcc}
      \toprule
      Age bucket & Male deaths & Female deaths \\
      \midrule    
$[0,4]$ & 410 & 325\\ 
$[5,9]$ & 27 & 27\\ 
$[10,14]$ & 41 & 38\\ 
$[15,19]$ & 115 & 68\\ 
$\cdots$ & $\cdots$ & $\cdots$ \\
$[75,79]$ & 12\ 730 & 9\ 202\\ 
$[80,84]$ & 15\ 125 & 12\ 899\\ 
$[85,89]$ & 14\ 737 & 17\ 246\\ 
$90+$ & 12\ 231 & 24\ 974\\
     \bottomrule
    \end{tabular}
    \caption{Male Dutch deaths in 2020. Eurostat weekly mortality data. \label{tab:firsthalf2020}}
\end{table}

\begin{figure}[h!]
\centering
\includegraphics[width = 0.8\textwidth]{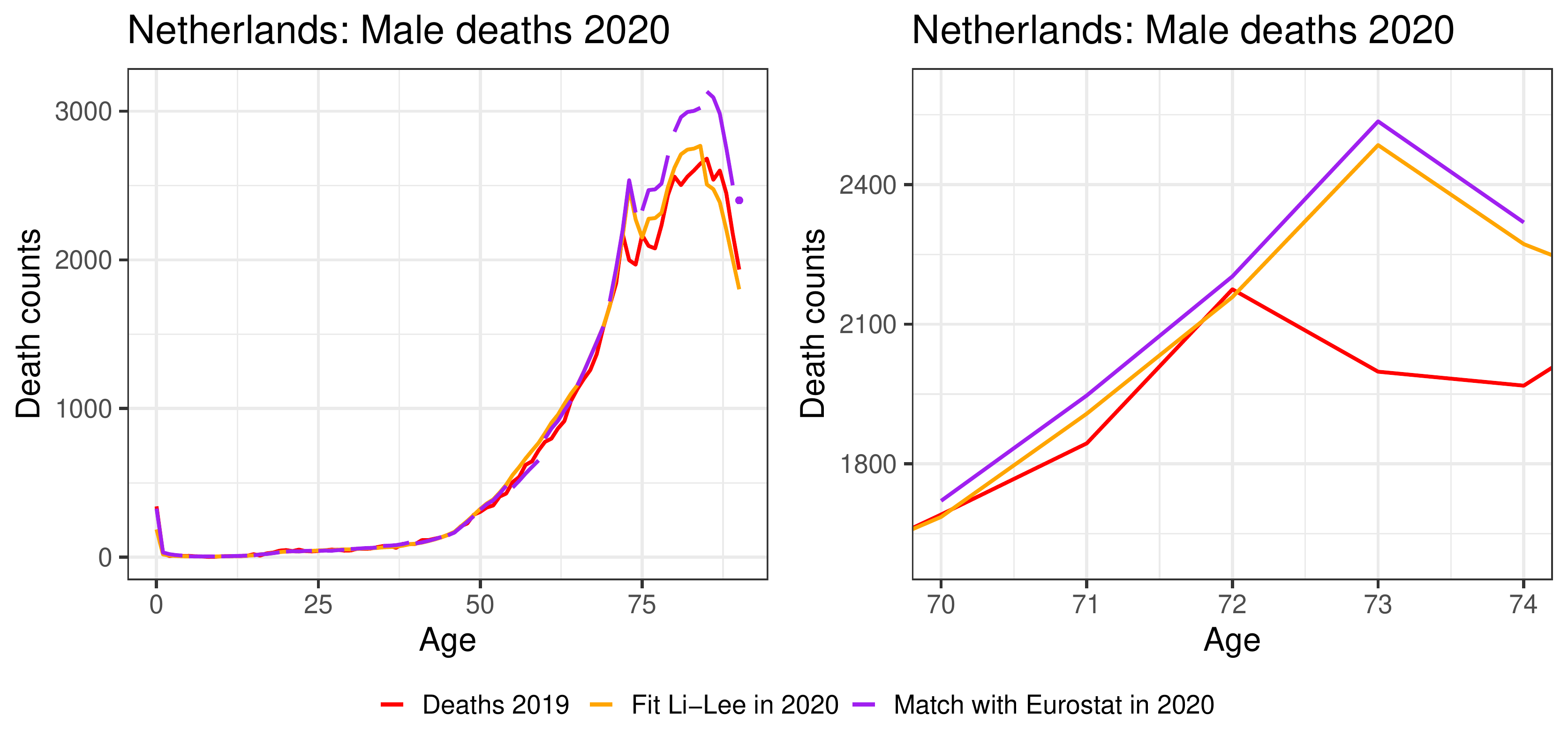}
\caption{Construction of the virtual deaths in individual ages in the Netherlands for 2020.\label{fig:virtdeathS1}}
\end{figure}

Having extracted the weekly deaths on Eurostat, we now return to the construction of the virtual deaths in $2020$ at individual ages 0-90. Figure~\ref{fig:virtdeathS1} graphically explains this construction for Dutch males. The red line shows the observed number of male deaths in the Netherlands for the year 2019 from the HMD. Applying the Li \& Lee mortality forecasting strategy of Section~\ref{subsec:futurepaths} to the Netherlands, we first project the force of mortality $\hat{\mu}_{x,t}^{\text{c}}$ for the year $t=2020$ and then calculate the estimated expected number of deaths $\hat{d}_{x,t}^{\text{c}}$. This corresponds to the orange line in Figure~\ref{fig:virtdeathS1}. Similar to the exposure matching principle in Appendix~\ref{sec:create.exp}, we then match the orange death curve of 2020 with the death counts collected in age buckets in 2020 from Eurostat, as given in Table~\ref{tab:firsthalf2020}. 

Denote $\hat{d}_{x,t}^s$ for the fitted death counts at age $x$ and time $t$ as obtained from the orange line in Figure~\ref{fig:virtdeathS1}. We consider an age bucket $[x_i,x_j]$ and define the virtual annual death counts $d_{x,t}$ as:
\begin{equation}\label{eq:deathstransition}
\begin{aligned}
d_{x,t} = \hat{d}_{x,t}^s \cdot b_{i,j}, \hspace{0.5cm} \text{where} \: \: \:
b_{i,j} = \dfrac{d_{[x_i,x_j],t}}{\displaystyle \sum_{a = x_i}^{x_j} \hat{d}_{a,t}^s},
\end{aligned}
\end{equation}
for $t = 2020$ and $x \in [x_i,x_j]$. Intuitively, we again vertically scale a section of the orange line, corresponding to a certain age bucket, such that the combined number of deaths within this age bucket corresponds to the total number of observed deaths in the same age bucket from Eurostat. This matching principle results in the purple line in Figure~\ref{fig:virtdeathS1}. The right panel of Figure~\ref{fig:virtdeathS1} illustrates the results for the age bucket $[70,74]$. This procedure leads to annual death counts $d_{x,t}$, now evaluated at individual ages.

The last age bucket $90+$ is an open age bucket. This implies that we have to modify our strategy outlined in Equation~\eqref{eq:deathstransition} to define the virtual death count at age $90$ in $2020$:
\begin{equation*}
\begin{aligned}
d_{90,2020} = d_{90,2018} + c_{90+}, \hspace{0.5cm} \text{where} \: \: \:
c_{90+} = A^g \left(d_{90+,2020} - \displaystyle \sum_{a = 90}^{\infty} d_{a,2018}\right),
\end{aligned}
\end{equation*}
and $A^g$ is a gender-specific rate, which we assume to be country-independent. For example, $A^g = 0.20$ means that $20\%$ of the deaths, at ages $90$ or higher, occur at age $90$. Based on the observed ratios of Belgium and Denmark in 2020,\footnote{For Belgium and Denmark, we already have the death counts at individual ages in 2020 from Statbel and HMD respectively. We take the average of both ratios.} we select $A^m = 0.2$ and $A^f = 0.145$ for males and females respectively.

We repeat this procedure for every European country in the study with the following country-specific data adjustments. For the United Kingdom, we do not have the deaths $d_{x,t}$ at time $t = 2019$ yet. In this case, we construct a Li \& Lee mortality model for the country of interest with a shorter calibration period, ending with the year $2018$. In addition, each country has its own starting year of the calibration period for stability reasons, e.g.~the year 1970 for the Netherlands. We can then construct death counts for the year $2019$ and $2020$ (for each scenario) by projecting the force of mortality for the years $2019$-$2020$ and by performing the matching principle at both years.

Moreover, for three of the European countries, namely Germany, France and the United Kingdom, we work with the weekly death counts in age buckets from the STMF data series, rather than from Eurostat.\footnote{We only use the weekly death counts collected in age buckets from Eurostat when they match the reported death counts in the larger age buckets from the STMF data series. We do this for safety reasons because some deviations between the weekly death counts on Eurostat and the STMF data series may occur due to for example territorial differences, e.g.~France with or without overseas regions.} For these countries we apply the strategy outlined above, although we use larger age buckets.

\end{document}